\documentclass[twocolumn,prb]{revtex4-1}

\usepackage{times}
\usepackage{graphicx}
\usepackage{amsfonts}
\usepackage{amsmath, amsthm, amssymb}
\usepackage{dsfont}
\usepackage{mathptm}
\usepackage{color}
\usepackage{subcaption}
\usepackage{tikz}
\usepackage{bm}
\usepackage[colorlinks=true, allcolors=black, citecolor=blue, linkcolor=blue]{hyperref}

\newcommand{\Imag}{{\Im\mathrm{m}}}   
\newcommand{\Real}{{\mathrm{Re}}}   
\newcommand{\ve}[1]{\mathbf{#1}}
\DeclareMathOperator{\diag}{diag} 

\newcommand{\vk}{\ve{k}} 
\newcommand{\Tr}{\mathrm{Tr}~}

\newcommand{\vs}{\bar{\bm{\sigma}}} 
\newcommand{\vp}{\ve{p}} 
\renewcommand{\vr}{\ve{r}} 

\newcommand{\vpf}{\mathbf{\vp}_\text{F}} 
\newcommand{\vpfu}{\hat{\ve{p}}_\text{F}} 
\newcommand{\vf}{v_\text{F}} 
 
\newcommand{\vq}{\ve{q}} 
\newcommand{\pa}[2][]{\psi_{#2}^{\vphantom{\dagger}#1}} 
\newcommand{\pc}[2][]{\psi_{#2}^{{\dagger}#1}}          
\newcommand{\su}{\uparrow}    
\newcommand{\sd}{\downarrow}  

\newcommand{\e}[1]{\mathrm{e}^{#1}}
\newcommand{\dif}{\mathrm{d}} 
\newcommand{\del}{\partial} 
\newcommand{\mean}[1]{\langle#1\rangle}
\newcommand{\Mean}[1]{\Big\langle#1\Big\rangle}

\newcommand{\paulimat}{\hat{\boldsymbol{\sigma}}}
\newcommand{\comm}[1]{\big[#1\big]}  
\newcommand{\anticomm}[1]{\big\{#1\big\}} 
\newcommand{\Comm}[1]{\Big[#1\Big]}  
\newcommand{\Anticomm}[1]{\Big\{#1\Big\}} 

\renewcommand{\t}{\underline{\tau}}
\newcommand{\s}{\bar{\sigma}}

\DeclareMathAlphabet\mathbfcal{OMS}{cmsy}{b}{n}

\newcommand{\ie}{\textit{i.e. }}
\newcommand{\eg}{\textit{e.g. }}
\newcommand{\etal}{\emph{et al.}}
\def\i{\mathrm{i}}                        

\newcommand{\g}{{\gamma}}
\newcommand{\gt}{\tilde{\gamma}}
\newcommand{\ga}{\acute{\gamma}}

\newcommand{\G}{\mathcal{G}}
\newcommand{\vG}{\mathbfcal{G}}
\newcommand{\uG}{\underline{\mathcal{G}}}
\newcommand{\F}{\mathcal{F}}
\newcommand{\uh}{\underline{h}}

\DeclareMathOperator{\starp}{\text{\raisebox{0.15ex}{\scalebox{0.65}{$\otimes$}}}} 
\DeclareMathOperator{\ringp}{\circ}
\DeclareMathOperator\arctanh{arctanh}

\usepackage{soul}

\begin{document}

\title{Quasiclassical theory for the superconducting proximity effect in Dirac materials}

\author{Henning G. Hugdal}
\author{Jacob Linder}
\author{Sol H. Jacobsen}
\email[Corresponding author: ]{sol.jacobsen@ntnu.no}
\affiliation{Department of Physics, NTNU, Norwegian University of
Science and Technology, N-7491 Trondheim, Norway}


\begin{abstract}
We derive the quasiclassical non-equilibrium Eilenberger and Usadel equations to first order in quantities small compared to the Fermi energy, valid for Dirac edge and surface electrons with spin-momentum locking $\ve{p}\cdot\vs$, as relevant for topological insulators. We discuss in detail several of the key technical points and assumptions of the derivation, and provide a Riccati-parametrization of the equations. Solving first the equilibrium equations for S/N and S/F bilayers and Josephson junctions, we study the superconducting proximity effect in Dirac materials. Similarly to related works, we find that the effect of an exchange field depends strongly on the direction of the field. Only components normal to the transport direction lead to attenuation of the Cooper pair wavefunction inside the F. Fields parallel to the transport direction lead to phase-shifts in the dependence on the superconducting phase difference for both the charge current and density of states in an S/F/S-junction. Moreover, we compute the differential conductance in S/N and S/F bilayers with an applied voltage bias, and determine the dependence on the length of the N and F regions and the exchange field. 
\end{abstract}

\maketitle

\section{Introduction}
The study of materials featuring symmetry-protected topological states has in recent years attracted much attention. Topological insulators represent a notable example of such systems, which are characterized by a topological invariant that is manifested physically \eg via the presence or absence of robust edge-states for thin-films (2D) or surface-states for bulk materials (3D) (see reviews~\cite{hasan_rmp_10,qi_rmp_11,wehling_aip_14}). Much of the exotic physics predicted to occur in topological insulators requires proximity to a superconducting host material, such as the appearance of Majorana zero modes.\cite{fu_prl_08} Therefore, it is of interest to establish a theoretical framework that is accurate, yet practical to work with analytically, and capable of treating superconducting order in Dirac materials not only in the idealized ballistic limit of transport, but also in the ``dirty", diffusive limit of frequent impurity scattering.

The quasiclassical theory of superconductivity \cite{eilenberger,larkin_jetp_69,usadel,eliashberg_jetp_72,larkin_jetp_75,shelankov_jltp_85} is a suitable candidate for describing the diffusive limit of topological insulators as it is known to account very well for phenomena such as the Josephson effect, bound-states, thermoelectric effects, and many more in conventional metallic hybrid structures.\cite{eschrig_rep_15, chandrasekhar_arxiv_04,belzig_sam_99,morten_03,rammer_rmp_86} Recently, this theory has also been expanded to incorporate the presence of strongly spin-polarized interfaces.\cite{eschrig_njp_15} The quasiclassical equations for topological insulator/superconductor structures with strong impurity scattering, \ie the Usadel equation, have recently been used in the study of the presence of vortices \cite{ioselevich_prb_12} and helical magnetization.\cite{zyuzin_arxiv_15} A quasiclassical treatment of the Dirac surface states of the topological He$^3$-B phase was given in Ref.~\onlinecite{wu_prb_13}. However, several basic features of the quasiclassical superconducting proximity effect in Dirac materials have not yet been studied in detail, such as the fundamental superconductor-normal and superconductor-ferromagnet bilayer stuctures. Moreover, an analysis of technical aspects such as how to parametrize the quasiclassical distribution functions that provide the kinetic equations out-of-equilibrium, and how to describe the full proximity effect regime with a numerically suitable Riccati-parametrization.\cite{schopohl_prb_95, schopohl_arxiv_98, jacobsen_prb_15}

Here, we address these issues and more by providing a detailed derivation of the Eilenberger and Usadel equations valid for generic Dirac materials with spin-momentum locking $\mathbf{p}\cdot\boldsymbol{\s}$ in the normal-state Hamiltonian. By considering a superconductor/normal (S/N) bilayer, a superconductor/ferromagnet (S/F) bilayer, and a superconductor/ferromagnet/superconductor (S/F/S) junction, we draw out features of the superconducting proximity effect that contrast with conventional metallic structures. We show that the effect of the exchange field on such systems depends greatly on the direction of the field,\cite{tanaka_prl_09, linder_prb_10} and we detail how this difference manifests in physical observables like the charge current and density of states. We also solve the non-equilibrium equations, calculating the differential conductance and electron distribution function in S/N and S/F bilayers. Our emphasis is on providing a detailed working of the derivation and an explanation of the underlying physical assumptions. We reproduce some existing experimental features of such systems, but are unable to reproduce certain previous theoretical predictions. More specifically, due to the assumption of the Fermi level $\mu$ being the largest energy scale in the system, and the resulting fixed spin-structure of the Green's function, we show that neither odd-frequency \cite{berizinskii} $s$-wave triplets nor a suppression of the $p$-wave component of the superconducting order (predicted to appear in topological insulator/superconductor structures in Ref. \onlinecite{yokoyama_prb_12} and Ref.~\onlinecite{tkachov_prb_13}, respectively) appear in our framework. We discuss how the spin locking due to the normalization condition for the Green's function plays a crucial role in terms of capturing these phenomena. The quasiclassical approach developed here could provide a useful framework to explore phenomena in superconducting spintronics \cite{linder_nphys_15} in the context of Dirac materials. 

The remainder of the article is organised as follows. In Sec.~\ref{sec:theory} we outline the quasiclassical theory and use it to provide the details of the derivation of the non-equilibrium Eilenberger equation for firstly edge, then surface Dirac electrons in the diffusive limit. We then discuss the analytical solutions to the equations in the weak proximity limit in Sec.~\ref{sec:results}, and present numerical results in the full proximity regime. We calculate physical observables and discuss the consequences of our findings. We conclude in Sec.~\ref{sec:conclusion} with a summary of the main points and broader impact of the results before providing a brief outlook for further developments. Further details of the key calculational steps are provided in the \hyperref[sec:appendix]{Appendix}.

\section{Theory}\label{sec:theory}
Dirac materials with spin-momentum locking can be described by a Hamiltonian resembling the relativistic Dirac Hamiltonian\cite{dirac} for massless fermions, written in second quantized form as
\begin{equation}
    H = -\i \vf\int \dif \vr \sum_{\alpha\beta} \pc{\alpha}(\vr)(\nabla - \i e\ve{A})\cdot \vs_{\alpha\beta}\pa{\beta}(\vr),
\end{equation}
where the Fermi velocity $\vf$ takes the role of the speed of light in vacuum, $e=-|e|$ is the electron charge, $\ve{A}$ the vector potential, $\vs=(\s^1, \s^2, \s^3)$ is the vector in coordinate space consisting of Pauli spin matrices in spin space, and $\pc{}(\ve{r})=(\pc{\su} \quad \pc{\sd})$ and $\pa{}(\ve{r})=(\pa{\su} \quad \pa{\sd})^{T}$ where $\pc{\su(\sd)}$ and $\pa{\su(\sd)}$ are the field operators creating or annihilating an electron with spin up (down) at position $\ve{r}$ respectively. The subscripts $\alpha,\beta =1,2$ are used to specify the spin space elements of $\pc{}(\ve{r})$, $\pa{}(\ve{r})$ and the matrix $\vs$. As this is the only term differing from the non-Dirac case, we treat only this term explicitly here, with the full Hamiltonian provided for reference in the \hyperref[sec:appendix]{Appendix}. In order to describe both equilibrium and non-equilibrium properties of Dirac materials with the above Hamiltonian, we will utilize the Keldysh Green's function technique to find an equation of motion for the Green's functions. We will use the notation $\check{G}$ for the full $8\times8$ Keldysh Green's function matrices, $\underline{G}$ for $2\times2$ matrices in particle-hole space, $\bar{G}$ for $2\times2$ matrices in spin space, and $\hat{G}$ for $4\times4$ matrices in particle-hole$\otimes$spin space, \ie $\hat{G} = \underline{G}\otimes \bar{G}$, where $\otimes$ denotes a Kronecker product. We also define the notation $\breve{G}$ for $4\times4$ matrices in Keldysh space where the spin structure is excluded, \ie $\check{G} = \breve{G}\otimes\bar{G}$.
The matrix structure thus allows straightforward identification of the respective subspaces to which they and any combinations belong, \ie the spin, Nambu (particle-hole) or Keldysh spaces, in-keeping with the literature on quasiclassical theory. However, later we will also explicitly decompose many of the matrices in order to expose the underlying structure in the subspaces.

\subsection{Quasiclassical approximations -- The Eilenberger equation}
The derivation of the Eilenberger equation for the Dirac case follows the the same steps as in the conventional case, see the \hyperref[sec:appendix]{Appendix} for further details. At sufficiently low temperatures, only electrons near the Fermi surface will take part in the dynamics of the system, giving $\ve{p}$ a pronounced peak at $\vpf$. \cite{chandrasekhar_arxiv_04,belzig_sam_99,morten_03} To make the substitution $\ve{p} \rightarrow \vpf$ we introduce the quasiclassical Green's function,
\begin{equation}
    \check{g}(\vr,t,\vpf,\epsilon) \equiv \frac{\i}{\pi}\int\dif \xi_p \check{G}(\vr,t,\ve{p},\epsilon),\label{quasiclassical_integral}
\end{equation}
where $\xi_p = \vf\ve{p}$, and the structure of the $8\times 8$ matrix in Keldysh space $\check{G}$ is given in the \hyperref[sec:appendix]{Appendix}. Since the Fermi wavelength $\lambda_F$ is much smaller than the superconducting correlation length $\xi_S$, we keep only terms to lowest order in $\nu = \lambda_F/\xi_S$. After performing the approximations known collectively as the quasiclassical approximations, we arrive at the Eilenberger equation in the Dirac case,
\begin{equation}
    \frac{\vf}{2}\anticomm{\nabla\check{g},\hat{\rho}^3\paulimat} = \i\comm{\epsilon\hat{\rho}^3+\vf e \ve{A}\cdot \paulimat-\vf\vpf\cdot\hat{\rho}^3\paulimat,\check{g}}^\circ,
\end{equation}
where we have defined the matrix $\hat{\rho}^3 = \mathrm{diag}(1,1,-1,-1)$, and the $\ringp$-product, $A\ringp B = \exp\{-\frac{\i}{2}(\del_t^A\del_\epsilon^B-\del_\epsilon^A\del_t^B)\} A B$.
As in the conventional case, we can now add the contributions from a superconducting pair potential, impurity and spin-flip scattering potentials and an exchange field. This results in
\begin{eqnarray}
    \frac{\vf}{2}\anticomm{\nabla\check{g},\hat{\rho}^3\paulimat}
    &&= \i\big[\epsilon\hat{\rho}^3 + \hat{\Delta} -\check{\Sigma}_\mathrm{imp} -\check{\Sigma}_\mathrm{sf} -\vf\vpf\cdot\hat{\rho}^3\paulimat \nonumber\\*
    &&+(\ve{h}+\vf e \ve{A})\cdot \paulimat ,\check{g}\big]^\circ,\label{Eilenberger}
\end{eqnarray}
where $\hat{\Delta} = i\t^1\otimes\s^2\Delta$ with the superconducting gap $\Delta$, for simplicity chosen to be real, $\ve{h}$ is the exchange energy, and $\check{\Sigma}_\mathrm{sf}$ and $\check{\Sigma}_\mathrm{imp}$ are the spin-flip and impurity scattering self-energies respectively.

In order to separate the particle-hole and spin space parts of the above anticommutator, we use the unitary transformation
\begin{equation}
    \hat{U} = \left(\begin{matrix}
        \s^0 & 0 \\
        0 & \s^2
    \end{matrix}\right).\label{Unitary_transf}
\end{equation}
This gives an Eilenberger equation in terms of the transformed Green's function $\check{\G} = \hat{U}\check{g}\hat{U}^\dagger$,
\begin{eqnarray}
    \frac{\vf}{2}\anticomm{\nabla\check{\G},\t^0\otimes\vs} &&= \i\comm{\epsilon\t^3\otimes\s^0+\i\Delta\t^1\otimes\s^0 - \check{\varsigma}_\mathrm{imp} - \check{\varsigma}_\mathrm{sf} \nonumber\\*
    &&+(\ve{h}+\vf e \ve{A})\cdot\t^3\otimes\vs -\vf\vpf\cdot\t^0\otimes\vs,\check{\G}}^\circ,\nonumber\\*\label{Eilenberger_transformed}
\end{eqnarray}
where we have defined $\check{\varsigma} \equiv \hat{U}\check{\Sigma}\hat{U}^\dagger$, and $\otimes$ here denotes the Kronecker product between the Pauli matrices $\t^i$ in particle-hole space and the Pauli matrices $\s^i$ in spin space. So far, our treatment is similar to Ref. \onlinecite{zyuzin_arxiv_15}.

\subsection{Normalization condition}
The above Eilenberger equation must be supplemented by a normalization condition. In the quasiclassical limit the Fermi energy $\epsilon_\mathrm{F} = \vf|\vpf|$ is by far the largest energy scale in the system, and consequently the dominating term in the commutator of Eq. (\ref{Eilenberger_transformed}) is $\vf\vpf\cdot\t^0\otimes\vs$. A bulk solution Green's function matrix to Eq. (\ref{Eilenberger_transformed}) must therefore commute with $\vf\vpf\cdot\t^0\otimes\vs$ to lowest order. In the non-superconducting state $\Delta=0$, possible solutions for the different matrices in particle-hole$\otimes$spin space obeying the necessary symmetries are \[\hat{\G}^{R/A} = \pm\t^3\otimes\left(\s^0 + \vpfu\cdot\vs\right)\] and \[\hat{\G}^K = 2\tanh(\epsilon/2k_\mathrm{B}T)\t^3\otimes(\s^0 + \vpfu\cdot\vs),\]
where the last result is calculated directly from Eq.~(\ref{deltG}) and the definition of $\hat{\G}^K$ using only the Fermi-Dirac distribution. Collecting the above in $\check{\G}$ and calculating $\check{\G}\check{\G}$ we get the possible normalization condition
\begin{equation}
    \check{\G}\circ\check{\G} = 2\t^0\otimes\left(\s^0 + \vpfu\cdot\vs\right).\label{normalization}
\end{equation}
In order to show that this is in fact a general normalization condition for Eq.~(\ref{Eilenberger_transformed}), the normalization in Eq.~(\ref{normalization}) must be consistent with this equation. This can be shown to be the case using a general parametrization for $\check{\G}$ \cite{bobkova_arxiv,bobkova}, where the dominant terms are
\begin{equation}
    \check{\G} =\breve{\G}'\otimes\s^0 + \breve{\G}''\otimes\vpfu\cdot\vs,\label{G_parametrization_ansatz}
\end{equation}
and further assuming $\breve{\G}' = \breve{\G}''$ (see the \hyperref[sec:appendix]{Appendix} for details.) Here $\breve{\G}'$ and $\breve{\G}''$ are functions in Keldysh space excluding the spin parts of the Green's functions. The reason these particular terms are dominant is that they commute with $\vf\vpf\cdot\t^0\otimes\vs$ in the Eilenberger equation, which in the quasiclassical limit is by far the largest term. The assumption $\breve{\G}'=\breve{\G}''$ means that the spin structure of $\hat{\G}^{R/A}$ is locked and proportional to the projector on to helical eigenstates.\cite{tkachov_prb_13}

We also notice that the parametrization
\begin{equation}
    \hat{\G}^K = \hat{\G}^R\circ\hat{h}-\hat{h}\circ\hat{\G}^A\label{GK_parametrization}
\end{equation}
solves the off-diagonal part of the normalization, $\hat{\G}^R\circ\hat{\G}^K+\hat{\G}^K\circ\hat{\G}^A=0$, when
\begin{equation}\label{eq:hpara}
\hat{h} = \underline{h}'\otimes\s^0 + \underline{h}''\otimes(\vpfu\cdot\vs).
\end{equation}
The reason $\hat{h}$ has to be parametrized this way is that the solution of $\hat{\G}^K$ has to commute with $\vpfu\cdot\vs$, meaning that $\hat{\G}^K$ has to have the same form as Eq.~(\ref{G_parametrization_ansatz}). In order for Eq.~(\ref{GK_parametrization}) to have this form, $\hat{h}$ must be parametrized as stated above. When inserting the parametrization for $\hat{h}$ into Eq.~(\ref{GK_parametrization}), we find that
\begin{equation}
    \hat{\G}^K = \left(\uG^R\ringp(\underline{h}'+\underline{h}'')-(\underline{h}'+\underline{h}'')\ringp\uG^A\right)\otimes(\s^0+\vpfu\cdot\vs),\label{h_param}
\end{equation}
and hence the spin structure of $\hat{\G}^K$ is locked in the same way as for $\hat{\G}^{R/A}$, allowing us to simplify Eq.~(\ref{G_parametrization_ansatz}) in the following way:
\begin{equation}
    \check{\G} =\breve{\G}\otimes(\s^0 + \vpfu\cdot\vs),\label{G_parametrization}
\end{equation}
where $\breve{\G}\ringp\breve{\G}=\t^0$, and $\breve{\G}$ has the same structure as the usual $8\times 8$ Keldysh Green's function, Eq.~(\ref{GK}). Equation (\ref{eq:hpara}) is a new result which establishes how the non-equilibrium distribution function matrix can be parametrized for quasiclassical Dirac materials. Note that the normalization condition is therefore generally valid for both equilibrium and non-equilibrium situations in superconductor-normal Dirac material structures, and it is a good approximation for doped and weakly magnetic topological insulators.

\subsection{Diffusive limit}
In the experimentally common ``dirty limit'', where the non-magnetic impurity scattering rate is high, we can expand the matrix $\breve{\G}$ in spherical harmonics,
\begin{equation}
    \breve{\G} \approx \breve{\G}_s + \vpfu\cdot\breve{\vG}_p.\label{spherical_harmonics}
\end{equation}
In this limit, we can also use the self-consistent Born approximation when manipulating the self-energies related to the impurity potentials. This leads to the following expressions for the self-energies (see Refs.~\onlinecite{morten_03,rammer_rmp_86}, and the \hyperref[sec:appendix]{Appendix} for details):
\begin{eqnarray}
    \check{\varsigma}_\mathrm{imp} &=& -\frac{\i}{2\tau}\mean{\check{\G}}_F,\label{imp_approx}\\
    \check{\varsigma}_\mathrm{sf} &=& -\frac{\i}{6\tau_\mathrm{sf}}\t^3\otimes\vs\mean{\check{\G}}_F\cdot\t^3\otimes\vs\label{sf_approx},
\end{eqnarray}
where $\mean{\cdot}_F$ denotes averaging over the Fermi surface.

\begin{figure}[htbp]
\includegraphics[width=0.89\columnwidth]{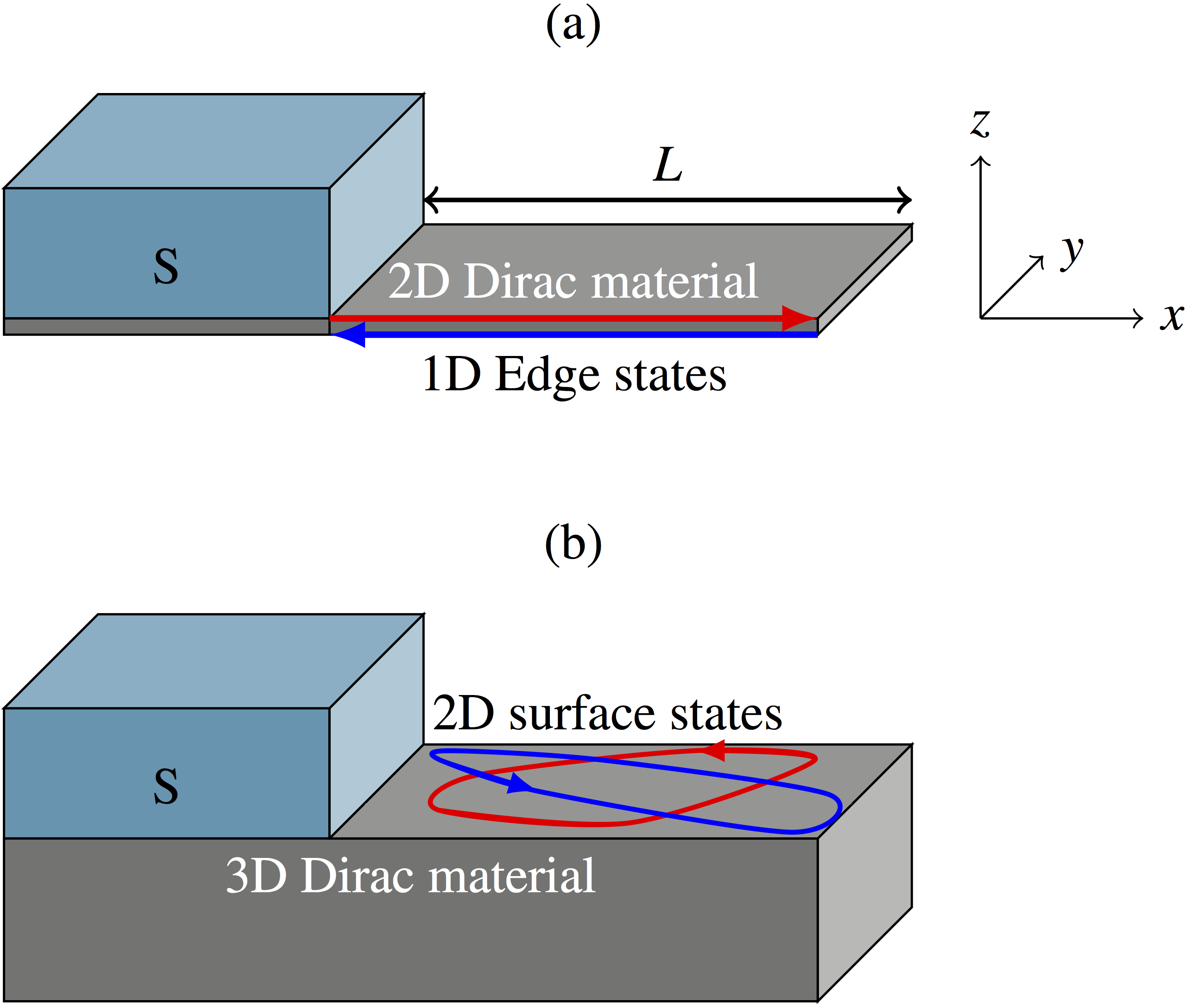}
\caption{\label{fig:model} Sketch of an SF junction in the two cases of (a) 2D Dirac materials with electron transport on a 1D edge, and (b) 3D Dirac materials with electron transport on a 2D surface. The figure also defines the coordinate system used.}
\end{figure}

When performing averages over the Fermi surface, the dimensionality of the Fermi surface greatly impacts the resulting equations. We hence treat the cases of electrons on a 1D edge (Fig.~\ref{fig:model}(a)) and 2D surface (Fig.~\ref{fig:model}(b)) separately.

\subsubsection{Dirac electrons moving on a 1D edge}
For simplicity we will drop the spin-flip term here.
Inserting the approximation Eq.~(\ref{imp_approx}) into the Eilenberger equation, Eq.~(\ref{Eilenberger_transformed}), together with the parametrization in Eq.~(\ref{G_parametrization}), and performing a trace over the spin space matrices yields
\begin{eqnarray}
       \ve{\vf}\cdot\nabla\breve{\G} &=& \i\Comm{\epsilon\t^3+\i\Delta\t^1+\frac{\i}{2\tau}\breve{\G}_s,\breve{\G}}^{\ringp}\nonumber \\*
       &+& \i\Comm{(\ve{h}+\vf e\ve{A})\t^3+\frac{\i}{2\tau}\breve{\vG}_p,\vpfu\breve{\G}}^{\ringp}.
\end{eqnarray}
In order to separate terms even and odd in $\vpfu$, we use Eq.~(\ref{spherical_harmonics}) and average over the Fermi surface after multiplying with the identity and $\vpfu$ respectively, the first case giving
\begin{equation}
    \vf\nabla\cdot\breve{\vG}_p = \i\Comm{\epsilon\t^3+\i\Delta\t^1,\breve{\G}_s}^{\ringp} +\i\comm{(\ve{h}+\vf e\ve{A}) \t^3,\breve{\vG}_p}^{\ringp},\label{Eilenberger_averaged}
\end{equation}
while first multiplying with $\vpfu$ and then averaging yields
\begin{equation}
    \vf\nabla\breve{\G}_s = \i\Comm{\epsilon\t^3+\i\Delta\t^1,\breve{\vG}_p}^\circ + \i\comm{(\ve{h}+\vf e\ve{A}) \t^3,\breve{\G}_s}^{\ringp}.
\end{equation}
We see that the impurity scattering term has completely dropped out from the above equations. This means that the usual method of using the fact that $\tau \to 0$ can not be used to express $\breve{\vG}_p$ in terms of $\breve{\G}_s$. However, exploiting the fact that we are considering a one-dimensional Fermi surface, we are able to combine the two above equations to a simplified equation for $\breve{\G}$,
\begin{equation}
    \ve{\vf}\cdot\hat{\nabla}\breve{\G} = \Comm{\i\epsilon\t^3 - \Delta\t^1,\breve{\G}}^{\ringp},\label{Usadel2D}
\end{equation}
where we have defined the operator
\begin{equation}
    \hat{\nabla} \breve{\G} = \nabla\breve{\G} -\frac{\i}{\vf}\comm{(\ve{h}+\vf e \ve{A})\t^3,\breve{\G}}^{\ringp}.\label{nabla_hat}
\end{equation}
We see that this is almost identical to the regular Eilenberger equation \cite{eilenberger} without any spin structure. The most prominent difference is that the exchange field $\ve{h}$ now enters the equation in the same way as the vector field $\ve{A}$. Note that with the assumptions made above, the $y$- and $z$-components of the fields $\ve{h}$ and $\ve{A}$ do not enter at all. The fact that the impurity scattering term does not enter the Eilenberger equation at all simply expresses that non-magnetic impurities cannot cause backscattering for Dirac electrons moving along an edge.

\subsubsection{Dirac electrons moving on a 2D surface}
We again expand the matrix $\breve{\G}$ in spherical harmonics, and assume that $|\breve{\vG}_p|\ll|\breve{\G}_s|$. Inserting this together with the expressions for the impurity self-energies, Eqs.~(\ref{imp_approx}) and (\ref{sf_approx}), into the transformed Eilenberger equation Eq.~(\ref{Eilenberger_transformed}), we get an equation with terms both even and odd in $\vpfu$. Again we separate the even and odd terms by averaging over the Fermi surface after multiplying with the identity and $\vpfu$ respectively, which yields the equations
\begin{equation}
    \vf\nabla\breve{\G}_s\otimes\vs + \frac{\vf}{2}\nabla(\breve{\vG}_p\cdot\vs)\otimes\vs = \i\Comm{\check{C}, \breve{\G}_s\otimes\s^0
    + \frac{1}{2}\breve{\vG}_p\otimes\vs}^{\ringp}\label{Eilenberger2D_averaged}
\end{equation}
and
\begin{eqnarray}
     \vf\nabla(\breve{\vG}_p\otimes\vs)\cdot\vs &+& \frac{\vf}{2}\anticomm{\nabla (\breve{\G}_s\otimes\vs_{\parallel}), \t^0\otimes\vs}\nonumber\\*
     &=& \i\Comm{\check{C}, \breve{\G}_s\otimes\vs_{\parallel}
    + \breve{\vG}_p\otimes\s^0}^{\ringp},\label{Eilenberger2D_pf_averaged}
\end{eqnarray}
where
\begin{eqnarray}
    \nonumber\check{C} &\equiv& \epsilon\t^3\otimes\s^0+\i\Delta\t^1\otimes\s^0
    + (\ve{h}+\vf e \ve{A})\cdot\t^3\otimes\vs \nonumber\\*
    &+& \frac{\i}{2\tau_{\mathrm{sf}}}\tau^3\big(\breve{\G}_s\otimes\s^0
    - \frac{1}{6}\breve{\vG}_p\otimes\vs\big)\t^3 +\frac{\i}{2\tau}\big(\breve{\G}_s\otimes\s^0
    + \frac{1}{2}\breve{\vG}_p\otimes\vs\big) \nonumber,
\end{eqnarray}
and the symbol $\parallel$ denotes that only the in-plane ($x$- and $y$-~) components of the vector enters the equation. Performing a trace over the spin-space matrices in Eqs.~(\ref{Eilenberger2D_averaged}) and (\ref{Eilenberger2D_pf_averaged}), neglecting terms second order in $\breve{\vG}_p$, we get
\begin{eqnarray}
    \frac{\vf}{2}\nabla\cdot\breve{\vG}_p &=& \i\Comm{\epsilon\t^3+i\Delta\t^1+\frac{\i}{2\tau_{\mathrm{sf}}}\t^3\breve{\G}_s\t^3,\breve{\G}_s}^{\ringp}\nonumber\\* &+& \frac{\i}{2}\comm{(\ve{h}_{\parallel}+\vf e \ve{A}_{\parallel})\t^3,\breve{\vG}_p}^{\ringp},\label{TrEilenberger2D_averaged}
\end{eqnarray}
and
\begin{eqnarray}
    \vf\nabla\breve{\G}_s &=& \i\Comm{\epsilon\t^3+\i\Delta\t^1+\frac{\i}{2\tau_{\mathrm{sf}}}\t^3\breve{\G}_s\t^3+\frac{\i}{4\tau}\breve{\G}_s,\breve{\vG}_p}^{\ringp}\nonumber\\* &+& \i\Comm{(\ve{h}_{\parallel}+\vf e \ve{A}_{\parallel})\t^3-\frac{\i}{12\tau_{\mathrm{sf}}}\t^3\breve{\vG}_p\t^3,\breve{\G}_s}^{\ringp}.\label{TrEilenberger2D_pf_averaged}
\end{eqnarray}


In the high-impurity limit, the mean time between scattering events $\tau$ becomes very small. Together with the assumption that $|\breve{\vG}_p|\ll|\breve{\G}_s|$, this allows us to neglect all terms linear in $\breve{\vG}$ except the impurity scattering term in Eq.~(\ref{TrEilenberger2D_pf_averaged}). Following the regular procedure \cite{usadel} to express $\breve{\vG}_p$ in terms of $\breve{\G}_s$, we arrive at
\begin{equation}
    \breve{\vG}_p = -2\tau\vf\breve{\G}_s\ringp\hat{\nabla}\breve{\G}_s,
\end{equation}
where $\hat{\nabla}$ is the operator defined in Eq.~(\ref{nabla_hat}) with only the in-plane components of the fields. Since $\nabla\breve{\G}_s\sim \epsilon\vf^{-1}$, where $\epsilon$ is small compared to the Fermi energy, we see that the assumption $|\breve{\vG}_p|\ll|\breve{\G}_s|$ holds in the high-impurity limit where $\tau\rightarrow 0$. Inserting the above into Eq.~(\ref{TrEilenberger2D_averaged}), and defining the diffusion constant $D\equiv \tau\vf^2/2$, we arrive at the Usadel equation for the isotropic matrix,
\begin{equation}
    2D\i\hat{\nabla}\cdot(\breve{\G}\ringp\hat{\nabla}\breve{\G}) = \left[\epsilon\t^3+\i\Delta\t^1+\frac{\i}{2\tau_{\mathrm{sf}}}\t^3\breve{\G}\t^3,\breve{\G}\right]^{\ringp},\label{Usadel3D}
\end{equation}
where we have dropped the subscript $s$. The form is very similar to the regular Usadel equation, with the significant difference that the exchange field enters in a way similar to the vector field.\cite{zyuzin_arxiv_15}

\subsection{Riccati parametrization}
The retarded component of Eq.~(\ref{Usadel3D}) can be solved numerically in the full proximity effect regime using the Riccati parametrization \cite{schopohl_prb_95,schopohl_arxiv_98,jacobsen_prb_15}. The particle-hole part of the transformed retarded Green's function matrix $\hat{\G}^R$ has the following symmetries,
\begin{equation}
    \uG^R = \left(\begin{matrix}
        \G & \F\\
        \tilde{\F} & -\G
    \end{matrix}\right),\label{uG_structure}
\end{equation}
where the $\tilde{(\cdot)}$ is the combined operation of complex conjugation and letting $\epsilon \to -\epsilon$. We have used that $\G = \tilde{\G}$, which follows from the normalization condition. This differs slightly from the symmetries of the regular retarded matrix $\hat{g}^R$ (due to the unitary transformation conducted initially), but using the regular parametrization in terms of the arbitrary unknown functions $\g$ and $\ga$ as an ansatz, $\G=N(1+\g\ga)$ and $\F=2N\g$ with $N=(1-\g\ga)^{-1}$, we find from the normalization for $\uG^R$ that $\tilde{\F} = -2N\ga$. Since $\F$ and $\tilde{\F}$ are related by the tilde operation, we must have $\ga = -\gt$. We therefore parametrize $\uG^R$ in the following way,
\begin{equation}
    \uG^R = N\left(\begin{matrix}(1-\g\gt) & 2\g\\
        2\gt & -(1-\g\gt)
    \end{matrix}\right),
\end{equation}
where $N=(1+\g\gt)^{-1}$. $\uG^R$ can thus be found by determining $\g$ and $\gt$. Since $\G$ and $\F$ are functions and not matrices, the Riccati parametrization is greatly simplified since $N$, $\g$ and $\gt$ all commute. Inserting this parametrization into \eqref{Usadel3D}, we get two differential equations for $\g$ and $\gt$:
\begin{widetext}
\begin{subequations}
\begin{eqnarray}
        D(\nabla^2\g - 2N\gt(\nabla\g)^2) &=& -\i\epsilon\g + \frac{\Delta}{2}(\g\g-1) + \frac{1}{\tau_{\mathrm{sf}}}\g(2N-1)
        + \frac{2\i D}{\vf}(\nabla\cdot\ve{h})\g + \frac{4\i D}{\vf}(2N-1)\ve{h}\cdot \nabla\g +\frac{4Dh^2}{\vf^2}(2N-1)\g,\\
        D(\nabla^2\gt - 2N\g(\nabla\gt)^2) &=& -\i\epsilon\gt + \frac{\Delta}{2}(\gt\gt-1) + \frac{1}{\tau_{\mathrm{sf}}}\gt(2N-1) -\frac{2\i D}{\vf}(\nabla\cdot\ve{h})\gt - \frac{4\i D}{\vf}(2N-1)\ve{h}\cdot \nabla\gt +\frac{4Dh^2}{\vf^2}(2N-1)\gt.
\end{eqnarray}
\end{subequations}
\end{widetext}
Notice that the second equation is the tilde-conjugate of the first. The above equations are a new result which renders a numerical treatment of the Usadel equation for Dirac materials particularly efficient.

\section{Results and discussion}\label{sec:results}
We derived the Eilenberger equation for a Dirac material Eq.~(\ref{Eilenberger}). Transforming this according to Eq.~(\ref{Unitary_transf}) and supplementing it with the normalization condition Eq.~(\ref{normalization}), we arrive at a simplified Eilenberger equation for the particle-hole part of the Keldysh matrix, Eq.~(\ref{Usadel2D}) supplemented by the normalization condition $\breve{\G}\ringp\breve{\G} = \t^0$.
For Dirac electrons moving on a surface, we have arrived at a Usadel equation for the isotropic particle-hole part of the Keldysh matrix, Eq.~(\ref{Usadel3D}).
In both the above equations we have defined the operator $\hat{\nabla}\G = \nabla\G - \i(\ve{h}_{\parallel}/\vf)\cdot\comm{\t^3,\G}^{\ringp}$, where the $\ringp$-product reduces to regular matrix multiplication in equilibrium. In the following, we first solve the retarded components of Eqs.~(\ref{Usadel2D}) and (\ref{Usadel3D}) for various systems in equilibrium. Afterwards, we proceed to solve the Keldysh-Usadel equation for the distribution function matrix $\uh$ (see the \hyperref[sec:appendix]{Appendix} for details), allowing us to study S/N and S/F structures brought out of equilibrium by an applied bias potential.

\subsection{Bulk solution in N, F and S}
Using Eqs.~(\ref{Usadel2D}) and (\ref{Usadel3D}), we find the bulk solution in a normal Dirac material (N) and proximity induced ferromagnet (F) to be $\uG^R=\t^3$. Adding the spin structure by Kronecker multiplying with the matrix $\s^0 + \vpfu\cdot\vs$ and using the unitary transformation $\hat{U}$ in Eq.~(\ref{Unitary_transf}), we find identical solutions for both N and F:
\begin{equation}
    \hat{g}^R_{N/F} = \left(\begin{matrix}
        \s^0 + \vpfu\cdot\vs & 0 \\
        0 & -\s^0 + \vpfu\cdot\vs^*
    \end{matrix}\right).
\end{equation}
We see that the above solution satisfies the necessary symmetries between the two diagonals. Furthermore, this solution is consistent with the fact that backscattering is supressed even in high-impurity Dirac materials due to the spin-momentum locking.\cite{wehling_aip_14,hasan_rmp_10} This can be seen by \eg calculating the spin-dependent density of states for Dirac edge electrons with spin in the $\pm x$-direction for $\vpfu = \pm\hat{x}$. Using the relations $\psi_{\su x}^\dagger = (\psi_{\su z}^\dagger+\psi_{\sd z}^\dagger)/\sqrt{2}$ and $\psi_{\sd x}^\dagger = (\psi_{\su z}^\dagger-\psi_{\sd z}^\dagger)/\sqrt{2}$ between the creation and annihilation field operators with spins in the $x$- and $z$-directions, we can express the density of states for spin up (down) in the $x$-direction as
\begin{equation}
    N_{\su(\sd)x} = \frac{N_0}{2}\Real\left\{\Tr[\bar{g}^R +(-) \bar{g}^R\s^1]\right\},
\end{equation}
where $N_0$ is the density of states per spin level in the normal state.
Using this result, we get $N_{\su x} = 2N_0$ and $N_{\sd x} = 0$ when $\vpfu = +\hat{x}$, and $N_{\su x} = 0$ and $N_{\sd x} = 2N_0$ when $\vpfu = -\hat{x}$, consistent with the fact that there is no backscattering from non-magnetic impurities.

The bulk solution in a proximity induced superconducting Dirac material is similar to that of a normal superconductor, the difference being the spin-structure of the resulting Green's function matrix:
\begin{eqnarray}
    \hat{g}^R_S = \left(\begin{matrix}
        c(\s^0 + \vpfu\cdot\vs) & s\e{\i\phi}(\s^0 + \vpfu\cdot\vs)\i\s^2\\
        s\e{-\i\phi}\i\s^2(\s^0 + \vpfu\cdot\vs) & -c(\s^0 - \vpfu\cdot\vs^*)
    \end{matrix}\right)
\end{eqnarray}
where we have used the $\theta$-parametrization,\cite{chandrasekhar_arxiv_04,belzig_sam_99} with $c= \cosh\theta$, $s=\sinh\theta$ and $\theta = \arctanh(|\Delta|/\epsilon)$. From the spin-structure, we see that both $s$- and $p$-wave pairing is present. We also see that a superconducting gap is present in the density of states, but with the possibility for a finite density of states only for electrons with spins in the direction of motion.

\subsection{Application: superconducting proximity effect for Dirac edge electrons in normal and ferromagnetic regions}
In the case of electrons on a 1D edge, the simplified Eilenberger equation Eq.~(\ref{Usadel2D}) can be solved exactly for both S/N and S/F structures. Using transparent boundaries, \ie continuity of the Green's functions, we find that the solution for the transformed retarded Green's function in an S/F structure is
\begin{equation}
    \uG^R(x) = \left(\begin{matrix}
        c & \i s \e{\frac{2\i}{\vf}(\epsilon p_x+h_x)x}\\
        \i s \e{-\frac{2\i}{\vf}(\epsilon p_x+h_x)x} & -c
    \end{matrix}\right),
\end{equation}
where $p_x=\pm 1$ depending on the direction of the Fermi momentum, and $h_x$ is the $x$-component of the exchange field. From the above we see that there is no attenuation of the anomalous components of $\uG^R$ due to the exchange field.

\subsection{Application: superconducting proximity effect for Dirac surface electrons in normal and ferromagnetic regions}
In the case of Dirac electrons moving on a surface, we solve the Usadel equation Eq.~(\ref{Usadel3D}) in the weak proximity regime. This regime is valid either when the interface transparency is low, or the system is close to the critical temperature, in both cases leading to weak superconducting correlations in the F region. Hence the solution in this case is expanded around the bulk solution $\t^3$, $\uG\approx \t^3 + \delta\uG$, where $\delta\uG$ has the matrix structure shown in Eq.~(\ref{uG_structure}). Using the normalization condition one can show that the corrections $\delta\G,~\delta\tilde{\G}$ to the normal components of $\uG$ is second order in the corrections $\delta\F,~\delta\tilde{\F}$ to the anomalous components, $\delta\G = - \delta\F\delta\tilde{\F}/2$. Hence we will focus on finding the solution only for $\delta\F$ and $\delta\tilde{\F}$ rather than all elements in $\delta\uG$. Assuming that the system varies only along the $x$-direction, we insert the expansion into Eq.~(\ref{Usadel3D}) and keep only terms to first order in $\delta\F,~\delta\tilde{\F}$. Solving the resulting differential equations we find the general solutions
\begin{subequations}
\begin{eqnarray}
    \delta\F &=& e^{\frac{2\i h_x x}{\vf}}(A_1\cosh kx + A_2\sinh kx),\\
    \delta\tilde{\F} &=& e^{-\frac{2\i h_x x}{\vf}}(A_3\cosh kx + A_4\sinh kx),
\end{eqnarray}
\end{subequations}
where $k = \sqrt{4h_y^2/\vf^2 - \i\epsilon/D}$, and $A_j, j=1,2,3,4$ are $x$-independent functions which must be determined by the boundary conditions.

In an S/F structure, where the solution must be equal to the superconducting bulk solution at the left boundary ($x=0$), and equal to zero at the right boundary ($x=L$), we get the following solution for the ferromagnetic region
\begin{equation}
    \uG_{F} = \t^3 - \left(\begin{matrix}
        0 & e^{\frac{2\i h_x x}{\vf}}\\
        e^{-\frac{2\i h_x x}{\vf}} & 0
    \end{matrix}\right)\i\sinh\theta \frac{\sinh k(x-L)}{\sinh kL}.
\end{equation}
From this we see that the Cooper pair correlation function oscillates and is damped in the F-region. Notice that the value of $h_x$ does not affect the penetration length of Cooper pairs into the F-region. However, increasing the exchange field in the $y$-direction increases the damping of the above functions, meaning that the Cooper pairs' penetration length into the F-region depends on only $h_y$, $\xi_F \sim |\vf/h_y|$, compared to $\sim \sqrt{D/|h|}$ in the normal case.\cite{buzdin_rmp_05} From this we see that the effect on the system differs greatly depending on whether the field points parallel or perpendicular to the transport direction.

In an S/F/S structure, where we assume that the absolute value of the superconducting gap is the same in the left and right superconductors, we arrive at the solution
\begin{widetext}
\begin{equation}
    \uG_{F} = \t^3
    - \left(\begin{matrix}
        0 & e^{\frac{2\i h_x x}{\vf}+\i\phi_L}\left[\sinh k(x-L)-e^{-\frac{2\i h_x L}{\vf}+\i\phi}\sinh kx\right]\\
        e^{-\frac{2\i h_x x}{\vf}-\i\phi_L}\left[\sinh k(x-L)-e^{\frac{2\i h_x L}{\vf}-\i\phi}\sinh kx\right] & 0
    \end{matrix}\right)\frac{\i\sinh\theta}{\sinh kL},
\end{equation}
\end{widetext}
where $\phi = \phi_R - \phi_L$ is the phase difference between the right and left superconductors, and $L$ is the length of the weak-link.

Starting from the expression for the probability current density in a Dirac material, $\ve{j}(\ve{r}) = \vf\sum_{\alpha\beta}\pc{\alpha}(\ve{r})\vs_{\alpha\beta}\pa{\beta}(\ve{r})$, we derive the following expression for the charge current density given in terms of the transformed Green's function matrices,
\begin{equation}
    \ve{j}_q = N_0e D \int \dif\epsilon \Tr\{\t^3(\breve{\G}\ringp\hat{\nabla}\breve{\G})^K\},
\end{equation}
where the superscript $K$ means that we take the Keldysh component of the matrix $\breve{\G}\ringp\hat{\nabla}\breve{\G}$. In equilibrium, this expression can be simplified to
\begin{equation}
    \ve{j}_q = N_0eD\int\dif\epsilon\Tr\{\t^3\uG^R\hat{\nabla}\uG^R+(\uG^{R}\hat{\nabla}\uG^{R})^{\dagger}\t^3\}\tanh\frac{\beta\epsilon}{2}.
\end{equation}
Inserting the above results for the S/F/S-junction, we arrive at the following expression for the current density in the $x$-direction,
\begin{eqnarray}
    j^x_q = &-&4N_0eD\sin\left(\phi - \frac{2h_xL}{\vf}\right)\nonumber\\* &\times&\int \dif\epsilon\Imag\left\{\frac{k}{\sinh kL}\right\}\sinh^2\theta\tanh\frac{\beta\epsilon}{2}.
\end{eqnarray}
\begin{figure}[h!]
\includegraphics[width=0.9\columnwidth]{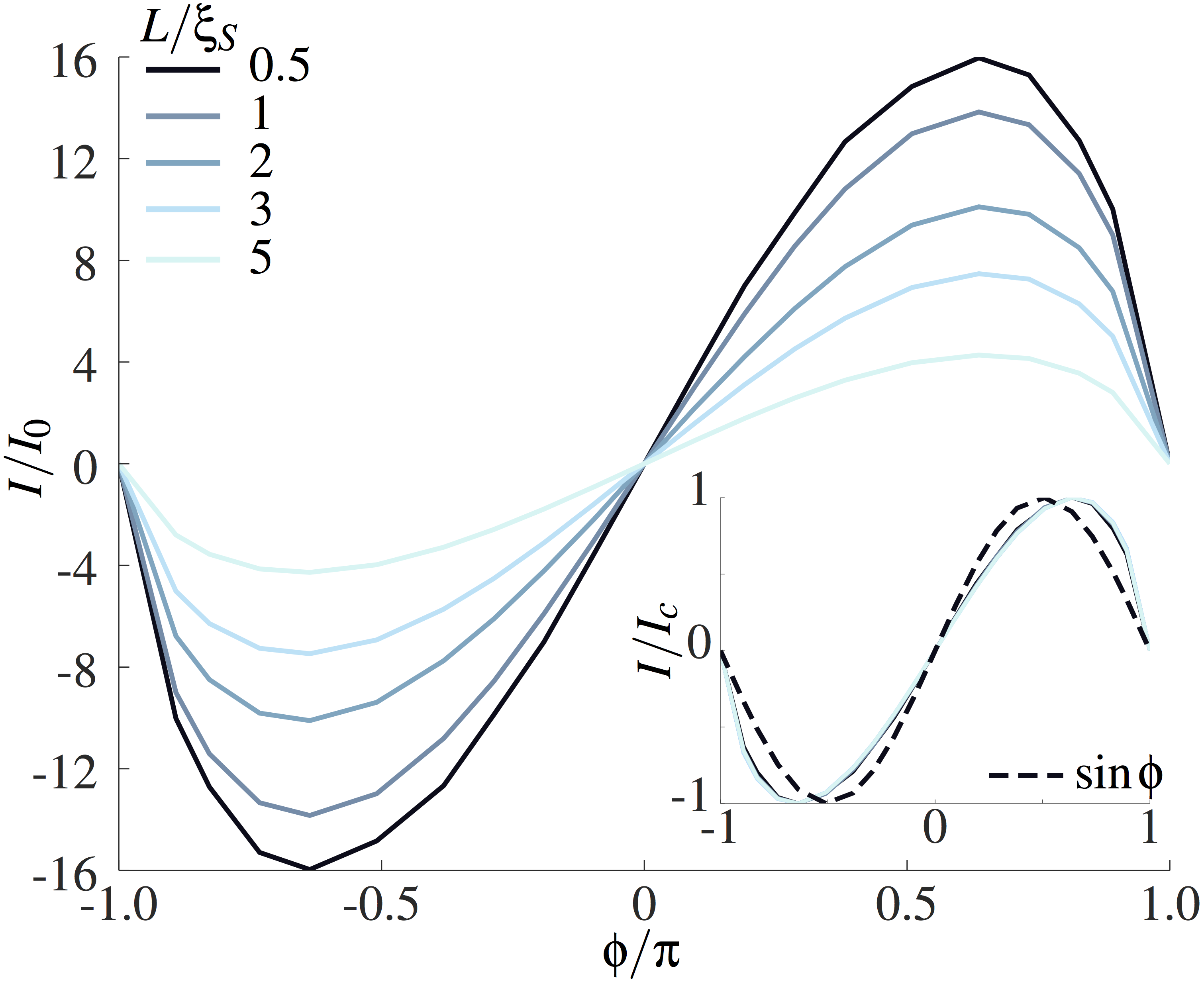}
\caption{\label{fig1}Current-phase relation of a Dirac S/N/S structure at temperature $T/T_c = 0.02$ for different lengths of the weak-link. The length is given in units of $\xi_S$, the diffusive coherence length of a bulk superconductor. The current decreases for increasing $L/\xi_S$, and is skewed compared to the regular $\sin\phi$-dependence, as shown in the inset where the current is normalized to the critical current.}
\end{figure}
We see that in the absence of exchange fields, the current follows the regular $\sin\phi$-dependence on the phase difference between the superconductors. In an S/F/S-junction, however, the $x$-component of the exchange field leads to a shift in the current-phase relation, consistent with previous findings.\cite{tanaka_prl_09} Since this shift depends on both $h_x$ and the length of the junction, the current at $\phi=0$ can in principle be tuned by the length of the weak-link.

A similar result is found when studying the local density of states.
The local spin-independent density of states is defined by $N(\epsilon, \vr) = N_0 \Real\{\Tr[\bar{g}(\epsilon,\vr)]\}/2$\cite{belzig_sam_99,chandrasekhar_arxiv_04}, where $N_0$ is the density of states per spin level at the Fermi level. Dividing by $N_0$ and switching to the transformed Green's functions, we find the normalized density of states
\begin{equation}
    D(\epsilon,\vr) = \Real\{\G(\epsilon,\vr)\} \approx 1 -\frac{1}{2}\Real\{\delta\F(\epsilon,\vr)\delta\tilde{\F}(\epsilon,\vr)\},
\end{equation}
where we have used the weak-proximity approximation to second order. Inserting the results for the S/F/S-junction, we get
\begin{eqnarray}
    D(\epsilon, x) = 1 &+& \Real\Bigg\{\frac{\sinh^2\theta}{2\sinh^2kL}\bigg[\sinh^2k(x-L) + \sinh^2kx \nonumber\\*
    &-& 2\sinh (k (x-L)) \sinh (kx)\cdot \cos\left(\phi - \frac{2h_x L}{\vf}\right)\bigg]\Bigg\}\nonumber.\\*\label{DOS_wp}
\end{eqnarray}

\begin{figure*}
\includegraphics[width=\textwidth]{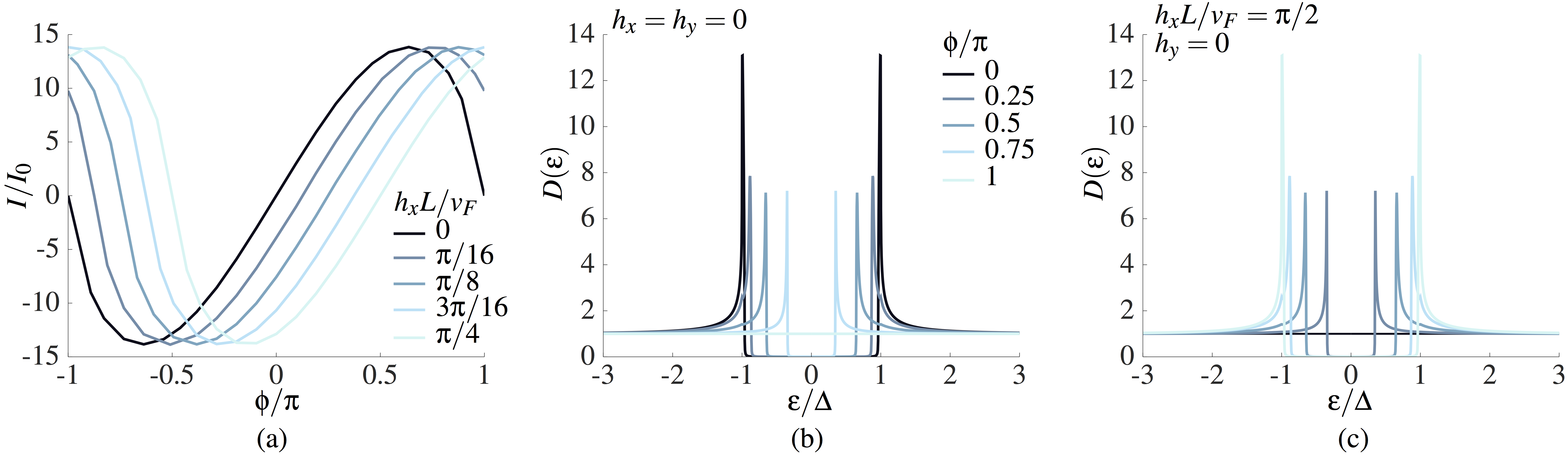}
\caption{\label{fig2}(a) Current-phase relation for different values of $h_xL/\vf$. We see that the $x$-component of the exchange field leads to a phase-shift $\delta = -2h_xL/\vf$, where $L$ is the length of the F-region. This is also the case for the normalized local density of states $D(\epsilon)$ at $x=L/2$, as seen when comparing the dependence on $\phi$ (b) with $h_x=0$, and (c) $h_xL/\vf=\pi/2$. In the latter case, the dependence on $\phi$ is inverted compared to the regular dependence. All results are obtained at temperature $T/T_c=0.02$.}
\end{figure*}

Motivated by the experiment of Sochnikov \etal \cite{sochnikov_prl_15}, we have solved the Usadel equation in the N-region of an S/N/S structure numerically using the Riccati parametrization and analyzed the full proximity effect. Calculating the current at temperatures close to the critical temperature, we get results in good correspondence with the analytical weak proximity results. At low temperatures, we get the current-phase relation shown in Fig.~\ref{fig1} for different junction lengths in the absence of an exchange field, where we have defined the constant  $I_0 = N_0eD\Delta A/L$. The figure shows that the current-phase relation is skewed compared to the regular $\sin\phi$-dependence, reproducing the experimental results reported in Ref. \onlinecite{sochnikov_prl_15}. Using $\epsilon_F = 0.05~\mathrm{eV}$ to estimate the density of states at the Fermi level, $N_0$, and parameter values from Sochnikov \etal, we find $I_0 = 0.2~\mathrm{{\mu}A}$. For \eg junction length $L=400~\mathrm{nm}$, corresponding to $L/\xi_S = 1$, we get a numerical value for the critical current, $I_C = 2.8\mathrm{~{\mu}A}$, which is in reasonable agreement with Ref. \onlinecite{sochnikov_prl_15}. Another interesting experimental finding was reported in Ref. \onlinecite{koren_epl_13}, who found signatures of induced triplet superconductivity in a superconductor/3D topological insulator bilayer.

The weak-proximity results showed that the $x$- and $y$-components of the exchange field affect the system in very different ways. This is also found to be the case when considering the full proximity effect: increasing $h_y$ lowers the critical current, while increasing $h_x$ leads only to a phase-shift $\delta=-2h_xL/\vf$ in the current-phase relation without changing the critical current, as shown in Fig.~\ref{fig2}(a). This is also found numerically to be the case for the density of states, where increasing $h_x$ affects the density of states only as a phase-shift in the $\phi$-dependence, in agreement with the weak-proximity results in Eq.~(\ref{DOS_wp}). At the value $h_xL/\vf=\pi/2$ the phase-dependence of the density of states is inverted compared to the normal case, as seen in Fig.~\ref{fig2}(b) and (c). Thus, with a finite $h_x$ the energy ground state of the system might also be shifted to a phase $\phi_0$ other than $0$ or $\pi$,\cite{buzdin_prl_08,grein_prl_09,kulagina_prb_14} where $\phi_0$ can be tuned by the exchange field and the length of the junction.  With a finite value of $h_y$, however, the density of states approaches that of a normal metal.

For various different values of the exchange field, we have not been able to produce a zero-energy peak in the DOS, a signature of odd-frequency spin-triplet pairing (although exceptions exist \cite{linder_scirep_15}), which has been theoretically predicted to be present in such structures.\cite{yokoyama_prb_12} Using the unitary transformation in Eq.~(\ref{Unitary_transf}) to transform solutions to Eq.~(\ref{Usadel3D}) back to the regular spin basis, we find the spin-structure of the anomalous matrix $\bar{f}$ to be
\begin{equation}
    \bar{f} = -i\F\left(\begin{matrix}
        -p_x+ip_y & 1\\
            -1 & p_x+ip_y
    \end{matrix}\right),
\end{equation}
where $\F$ is the particle-hole part of the solution, and $p_x$ and $p_y$ are the components of the unit vector $\vpfu$. First of all, we notice that the solution has both spin-singlet and -triplet components. However, due to the factors of $p_x,~p_y$, the spin-triplet components have $p$-wave pairing, not odd-frequency $s$-wave pairing. Getting spin-triplet solutions with odd-frequency $s$-wave pairing would require the introduction of a factor of both $\epsilon$ and $p_x$ or $p_y$. However, the latter is impossible from the $\vpfu$-averaged Usadel equation, and it therefore seems that it is not possible to get solutions including $s$-wave odd-frequency triplet Cooper pairs from the Usadel equation describing Dirac electrons moving on a surface.

The presence of $p$-wave pairing in proximity induced superconducting TIs has been found theoretically also in \eg Refs.~\onlinecite{yokoyama_prb_12,stanescu_prb_10,blackschaffer_prb_11}, while it has been theoretically predicted that the $p$-wave component is suppressed compared to the $s$-wave component in a disordered TI.\cite{tkachov_prb_13} Since both components are described by the same particle-hole function in our solution, there can be no such suppression of the $p$-wave component using this model.

The reason for the lack of odd-frequency $s$-wave components and the lack of a suppression of the $p$-wave component is the imposition of spin-locking by the assumption $\breve{\G}' = \breve{\G}''$ done when proving the normalization condition Eq.~(\ref{normalization}). In neglecting terms on the grounds that the Fermi energy is by far the largest energy scale in the system, we lose the possibility of changing the spin-structure of the Green's functions. The implications of this assumption with regard to \eg the absence of odd-frequency correlations has not been discussed in previous works.\cite{zyuzin_arxiv_15} We note that our results are consistent with Ref. \onlinecite{yokoyama_prb_12} in the quasiclassical limit $\mu \gg h$, since the odd-frequency amplitude is smaller than the even-frequency one by a factor $(h/\mu)^2$. Nevertheless, further work towards equations obtained when keeping terms small compared to $\breve{\G}'$, $\breve{\G}''$ in the parametrization Eq.~(\ref{G_parametrization_ansatz}) would be necessary in attempting to resolve the different predictions.

\begin{figure*}[t!]
    \includegraphics[width=\textwidth]{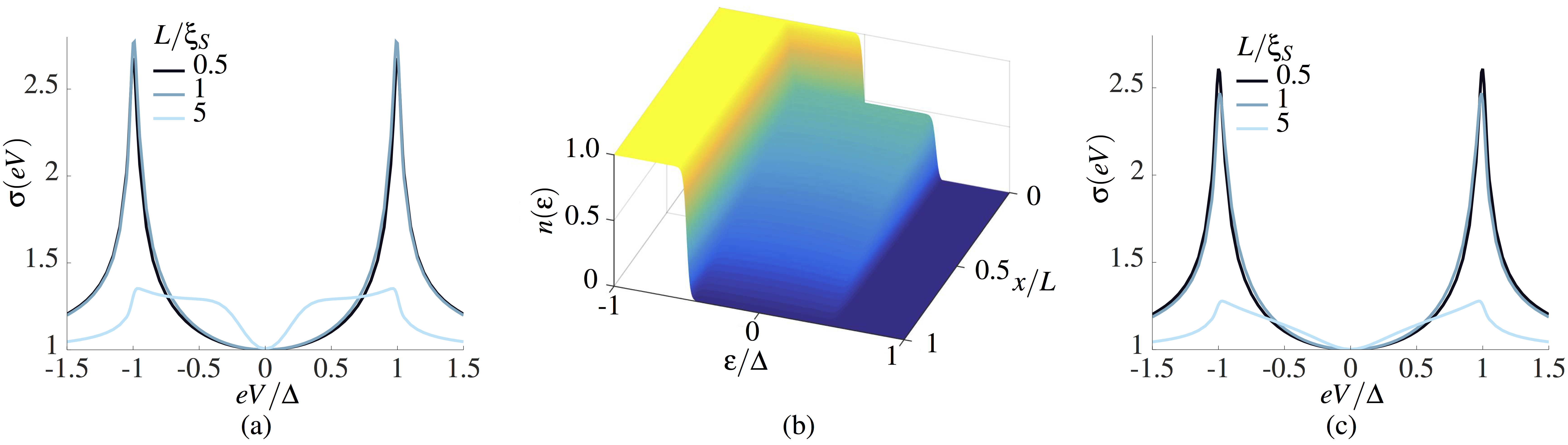}
    \caption{\label{fig3} (a) Normalized differential conductance in an S/N structure with potential bias $V$ for different lengths of the N-region. (b) Distribution function $n(\epsilon)$ for electrons in the N-region of an SN bilayer with potential bias $eV/\Delta=0.5$ applied to the boundary at $x/L=1$. The length of the N-region is $L/\xi_S = 1$, and $T/T_c = 0.02$. (c) Normalized differential conductance in an S/F structure with $h_x=0$, $h_y/\Delta=10$.}
\end{figure*}


\subsection{Application: proximity effect in non-equilibrium normal and ferromagnetic regions}
In order to study non-equilibrium systems, we solve the Keldysh component of the Usadel equation Eq.~(\ref{Usadel3D}) using the parametrization in terms of the matrix $\underline{h}$ [Eq.~(\ref{h_param})] for S/N and S/F bilayers with a potential bias $V$ applied to the boundary at $x/L = 1$. Using the parametrization $\underline{h} = h_L\t^0+h_T\t^3$ this amounts to solving two uncoupled equations for $h_L$ and $h_T$ using the solutions for the retarded and advanced Green's functions (see the \hyperref[sec:appendix]{Appendix} for details). For an S/N structure, the differential conductance $\sigma = \dif I/\dif V$ (normalized against its normal-state value obtained at $eV\gg\Delta$), shown for different lengths of the normal region in Fig.~\ref{fig3}a, displays behaviour similar to the non-Dirac case.\cite{tanaka_prb_03} The distribution function for electrons, defined by $n = (1-h_L-h_T)/2$, at potential bias $eV/\Delta = 0.5$ is shown in Fig.~\ref{fig3}b. This differs from the non-equilibrium N/N case in that the step in Fig.~\ref{fig3}b has twice the width but only half the height compared to the N/N case.\cite{chandrasekhar_arxiv_04}

\begin{figure}
    \includegraphics[width=0.95\columnwidth]{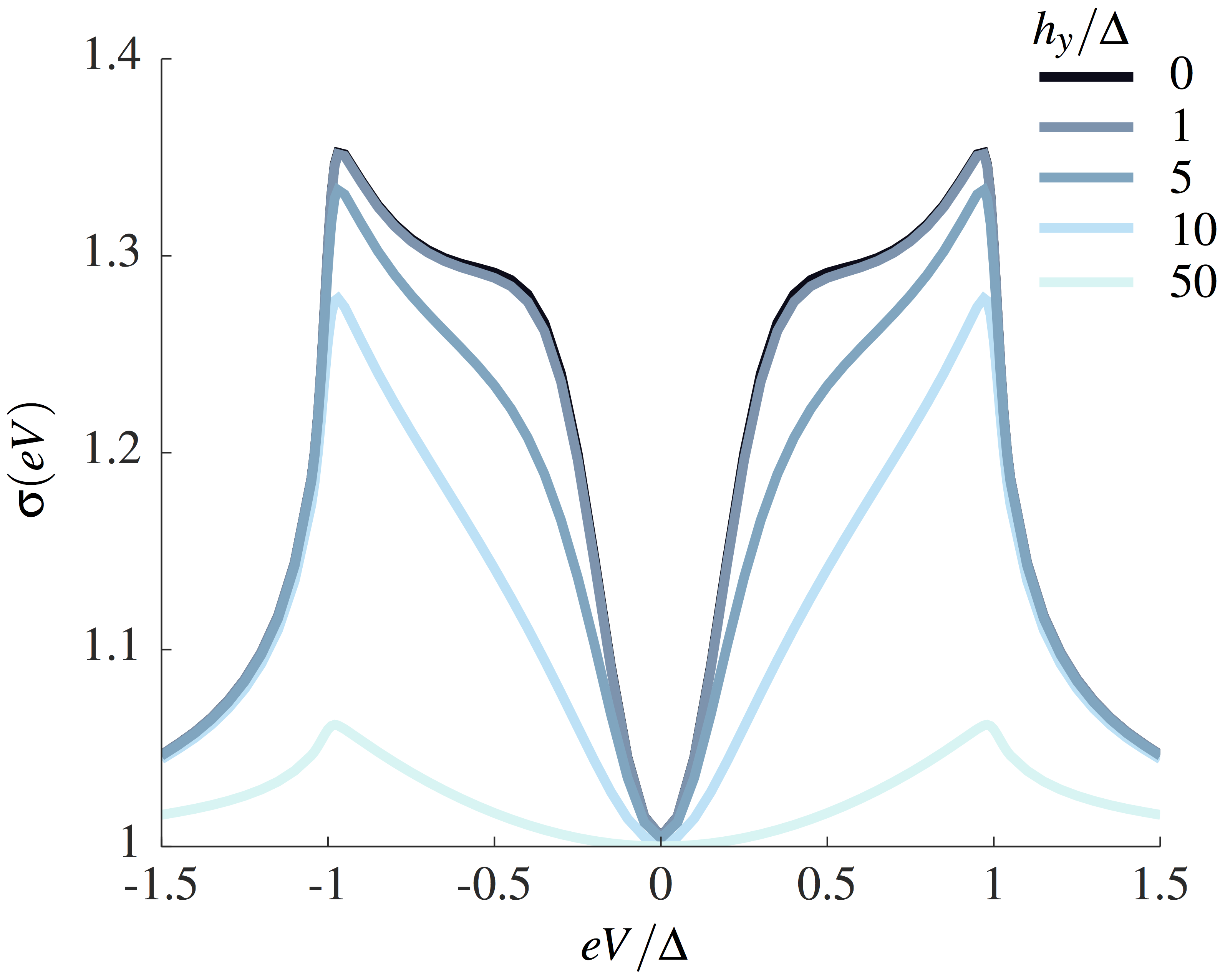}
    \caption{\label{fig4} Normalized differential conductance in S/F structure with $L/\xi_S=5$ at different values of $h_y/\Delta$. As $h_y$ is increased, both the peaks at $|eV|/\Delta = 1$ and the low-bias conductance is suppressed, approaching that of a N/N structure at high $h_y$.}
\end{figure}

Including an exchange field in the $x$-direction does not alter the above result, since the field neither changes the solution of the retarded and advanced Green's functions, nor directly enters the transport equations for the elements of $\uh$. However, increasing $h_y$ does affect the solution, as shown in Fig.~\ref{fig3}c for $h_y/\Delta = 10$. We see that increasing $h_y$ leads to a small reduction of the peaks of $\sigma$ around $eV/\Delta=1$, and a suppression of the low-bias conductance feature for longer sample lengths. This is further highlighted in Fig.~\ref{fig4}, where we plot the differential conductance for different values of $h_y$. When increasing the exchange field, the peaks at $eV/\Delta=\pm1$ and the low-bias conductance is suppressed, approaching that of a N/N structure ($\sigma = 1$) at high $h_y$. This is consistent with the fact that the superconducting correlations in the F-region are suppressed when increasing the exchange field in the $y$-direction.



\section{Concluding remarks}\label{sec:conclusion}
In summary, we have derived the quasiclassical non-equilibrium Eilenberger and Usadel equations for Dirac edge and surface electrons with spin-momentum locking. By studying S/N, S/N/S, S/F and S/F/S structures, we have shown that both singlet $s$-wave and triplet $p$-wave superconductivity is induced in the normal and ferromagnetic regions. Moreover, we have shown that the different directions of the exchange field affect the systems in significantly different ways, the penetration length of Cooper pairs into the F-region depending only on the fields perpendicular to the transport direction, $\xi_F\sim |\vf/\ve{h}_{\perp}|$. This difference is also clearly seen in the results for the density of states and charge current in an S/F/S-junction, where the exchange field in the transport direction leads to a phase shift.\cite{tanaka_prl_09} We have also shown that the charge current for an S/N/S-junction is skewed compared to the regular $\sin\phi$-dependence, in agreement with experimental results.\cite{sochnikov_prl_15} Moreover, we have found results for the differential conductivity which resemble the non-Dirac case for S/N stuctures with a potential bias,\cite{tanaka_prb_03} and showed how these results are changed by increasing the $y$-component of the exchange field.

An important purpose of our work has been to provide an in-depth analysis of technical aspects such as how to parametrize the quasiclassical distribution functions that provide the kinetic equations out-of-equilibrium, and how to describe the full proximity effect regime with a numerically suitable Ricatti-parametrization.\cite{schopohl_prb_95, schopohl_arxiv_98} Due to the approximations made during the derivation, keeping only lowest order terms, we have not been able to find signatures of odd-frequency $s$-wave pairing \cite{yokoyama_prb_12} or suppression of the $p$-wave component of the superconducting order parameter.\cite{tkachov_prb_13} This problem might be solved by keeping terms small compared to $\breve{\G}'$, $\breve{\G}''$ in the parametrization Eq.~(\ref{G_parametrization_ansatz}), in this way avoiding the spin-locking of the Green's functions. Further work is also needed in deriving more general boundary conditions valid in Dirac materials, since the spin-momentum locking has consequences when introducing boundaries between different materials.

\begin{acknowledgments}
We thank I. V. Bobkova and A. M. Bobkov for helpful discussions.
J.L was supported by the Research Council of Norway, Grants No. 205591, 216700, 240806 and the "Outstanding Academic Fellows" programme at NTNU.
\end{acknowledgments}

\appendix*

\section{Details of the derivation}\label{sec:appendix}
The full system Hamiltonian reads
\begin{eqnarray}
    \nonumber H = &-&\i \vf\int \dif \vr \sum_{\alpha\beta} \pc{\alpha}(\vr)(\nabla - \i e\ve{A})\cdot \vs_{\alpha\beta}\pa{\beta}(\vr)\nonumber\\*
&+&\int\dif\vr~\left(\Delta(\vr)\pc{\su}(\vr)\pc{\sd}(\vr)+\Delta^\dagger(\vr)\pa{\sd}(\vr)\pa{\su}(\vr)\right)\nonumber\\*
    &+&\int\dif\vr\sum_{\alpha} V_\mathrm{imp}(\vr)\pc{\alpha}(\vr)\pa{\alpha}(\vr)\nonumber\\*
    &+& \int\dif\vr\sum_{\alpha\beta}\pc{\alpha}(\vr)V_\mathrm{sf}\ve{s}(\vr) \cdot \vs_{\alpha\beta}\pa{\beta}(\vr)\nonumber\\*
    &-& \int\dif\vr\sum_{\alpha\beta}\pc{\alpha}(\vr)\ve{h}(\vr) \cdot \vs_{\alpha\beta}\pa{\beta}(\vr),\label{system_hamiltonian}
\end{eqnarray}
where $\Delta$ is the superconducting pair potential, $V_\mathrm{imp}$ the impurity potential, $V_\mathrm{sf}\ve{s}$ the spin-flip potential, and $\ve{h}$ the exchange field. Since only the kinetic term differs from the non-Dirac case, we include only this term the following derivation.

The normal and anomalous retarded (R), advanced (A) and Keldysh (K) Green's functions are defined by
\begin{eqnarray*}
    G_{\alpha\beta}^R(\vr,t;\vr',t') &=& -\i\mean{\anticomm{\pa{\alpha}(\vr,t),\pc{\beta}(\vr',t')}}\theta(t-t'),\\
    F_{\alpha\beta}^R(\vr,t;\vr',t') &=& -\i\mean{\anticomm{\pa{\alpha}(\vr,t),\pa{\beta}(\vr',t')}}\theta(t-t'),\\
    G_{\alpha\beta}^A(\vr,t;\vr',t') &=& +\i\mean{\anticomm{\pa{\alpha}(\vr,t),\pc{\beta}(\vr',t')}}\theta(t'-t),\\
    F_{\alpha\beta}^A(\vr,t;\vr',t') &=& +\i\mean{\anticomm{\pa{\alpha}(\vr,t),\pa{\beta}(\vr',t')}}\theta(t'-t),\\
    G_{\alpha\beta}^K(\vr,t;\vr',t') &=& -\i\mean{\comm{\pa{\alpha}(\vr,t),\pc{\beta}(\vr',t')}},\\
    F_{\alpha\beta}^K(\vr,t;\vr',t') &=& -\i\mean{\comm{\pa{\alpha}(\vr,t),\pa{\beta}(\vr',t')}}.
\end{eqnarray*}
Using the Heisenberg equation of motion for an operator $\mathcal{O}$, $i\del_t \mathcal{O} = \comm{\mathcal{O},H}$, we find the time-derivatives of the field operators. This in turn can be used to find the $t$ and $t'$-derivatives of the above Green's functions, which are collected in the following way,
\begin{subequations}
\label{deltG}
\begin{eqnarray}
    i\hat{\rho}^3\del_t \check{G} &=& \delta(t-t')\delta(\vr-\vr') + \hat{K}\check{G},\label{delt}\\
    -i\del_{t'} \check{G}\hat{\rho}^3 &=& \delta(t-t')\delta(\vr-\vr') + \check{G}\hat{K}'^\dagger.\label{deltprime}
\end{eqnarray}
\end{subequations}
Here we have defined the $8\times8$-matrix Keldysh space,
\begin{equation}
        \check{G} = \left(\begin{matrix}
            \hat{G}^R & \hat{G}^K\\
            0 & \hat{G}^A
        \end{matrix}\right)\label{GK},
\end{equation}
consisting of the $4\times4$-matrices in particle-hole$\otimes$spin space,
\begin{subequations}
\begin{eqnarray}
        \hat{G}^{R/A} = \left(\begin{matrix}
            \bar{G}^{R/A} & \bar{F}^{R/A} \\
            \bar{F}^{R/A*} & \bar{G}^{R/A*}
        \end{matrix}\right),\\*
        \hat{G}^{K} = \left(\begin{matrix}
            \bar{G}^{K} & \bar{F}^{K} \\
            -\bar{F}^{K*} & -\bar{G}^{K*}
        \end{matrix}\right),
\end{eqnarray}
\end{subequations}
and the $2\times2$ Green's function matrices in spin space,
\begin{subequations}
\begin{eqnarray}
            \bar{G}^{R/A/K} = \left(\begin{matrix}
                G^{R/A/K}_{\su\su} & G^{R/A/K}_{\su\sd} \\
                G^{R/A/K}_{\sd\su} & G^{R/A/K}_{\sd\sd}
            \end{matrix}\right),\\*
            \bar{F}^{R/A/K} = \left(\begin{matrix}
                F^{R/A/K}_{\su\su} & F^{R/A/K}_{\su\sd} \\
                F^{R/A/K}_{\sd\su} & F^{R/A/K}_{\sd\sd}
            \end{matrix}\right).
\end{eqnarray}
\end{subequations}
In addition we have defined the matrices $\hat{\rho}^3 \equiv \t^3\otimes\s^0=\diag(1,1,-1,-1)$, $\bar{K} = -\i\vf(\nabla - \i e\ve{A})\cdot \vs$, and $\hat{K} = \diag(\bar{K}, \bar{K}^*)$ for notational simplicity, where $\nabla$ acts to the left or right according to the matrix with which it is multiplied. Note that Kronecker products with identity matrices are implied to resolve products between matrices of different dimensions in Eq.~(\ref{deltG}). In addition, a prime (\eg $\hat{K}'$) denotes that the matrix function is a function of the primed coordinates $\vr'$ and $t'$. Subtracting Eq.~(\ref{deltprime}) from Eq.~(\ref{delt}) yields
\begin{eqnarray}
    \i \hat{\rho}^3\del_t\check{G}(\vr,t;\vr',t') &+& \i \del_{t'}\check{G}(\vr,t;\vr',t')\hat{\rho}^3 \label{delt_subtracted}\\* 
    &=&\hat{K}\check{G}(\vr,t;\vr',t')-\check{G}(\vr,t;\vr',t')\hat{K}'^{\dagger}.\nonumber
\end{eqnarray}
Since we are interested in the two-particle wave functions describing superconductivity, we perform a coordinate transformation to the mixed representation, expressing the above equation in terms of the center-of-mass coordinates $\vr_\mathrm{COM} = (\vr+\vr')/2$ and $T = (t+t')/2$, and the relative coordinates $\vr_\mathrm{rel} = \vr-\vr'$ and $\tau = t-t'$. Fourier transforming with respect to the relative variables, the above equation can be expressed as
\begin{eqnarray}
    \comm{\epsilon\hat{\rho}^3, \check{G}(\vr,t,\ve{p},\epsilon)}^{\starp} &=& -\frac{\i\vf}{2}\anticomm{\nabla\check{G}(\vr,t,\ve{p}, \epsilon),\hat{\rho}^3\paulimat}\nonumber\\* 
    &+& \vf \comm{\ve{p}\cdot\hat{\rho}^3\paulimat, \check{G}(\vr,t,\ve{p},\epsilon)}\nonumber\\*
    &-& \vf e \comm{\ve{A}\cdot\paulimat, \check{G}(\vr,t,\ve{p},\epsilon)}^{\starp},\label{EOM_fourier}
\end{eqnarray}
where we have defined the matrix $\paulimat = \diag(\vs, \vs^*)$, and let $\vr_\textrm{COM}\rightarrow \vr$, $T \rightarrow t$. The symbol $\starp$ in the superscript denotes a convolution over the variables,\cite{belzig_sam_99} which can be expressed as
\begin{eqnarray}
    A \starp B &=& e^{\frac{\i}{2}(\nabla_r^A\nabla_p^B-\nabla_p^A\nabla_r^B)} A\circ B \nonumber \\*
    &=& e^{\frac{\i}{2}(\nabla_r^A\nabla_p^B-\nabla_p^A\nabla_r^B)}
    e^{-\frac{\i}{2}(\del_t^A\del_\epsilon^B-\del_\epsilon^A\del_t^B)}
    A B, 
\end{eqnarray}
which also defines the $\ringp$-product. Note that a dot product between $\nabla$ and $\paulimat$ is implied in the first term on the right hand side of Eq.~(\ref{EOM_fourier}). Moreover, since $\vf\ve{p}$ has no explicit time-dependence, we can write the second commutator on the right hand side of Eq.~(\ref{EOM_fourier}) as a $\ringp$-commutator. Performing the quasiclassical approximations, including the additional terms and self-energies (see the next section) from the full system Hamiltonian in Eq.~(\ref{system_hamiltonian}), we arrive at the Eilenberger equation, Eq.~(\ref{Eilenberger}).

\subsection{Self-consistent Born approximation}
The impurity potentials can be treated using the self-consistent Born approximation. We use the Dyson equation \cite{rammer_rmp_86}
\begin{eqnarray}
    \delta(t-t')\delta(\vr-\vr') &=& (\i\hat{\rho}^3\del_t - \hat{K})\check{G}(\vr, t; \vr', t')\\* 
    &-& \int\dif \vr'' \int\dif t'' \check{\Sigma}(\vr, t; \vr'', t'')\check{G}(\vr'', t''; \vr', t') \nonumber,
\end{eqnarray}
to incorporate the impurity potentials via a self-energy term. The conjugate equation with $t'$ and $\vr'$-derivatives reads
\begin{eqnarray}
    \delta(t-t')\delta(\vr-\vr') &=& \check{G}(\vr, t; \vr', t')(\i\hat{\rho}^3\del_t' - \hat{K}') \\*
    &-& \int\dif \vr'' \int\dif t''\check{G}(\vr, t; \vr'', t'') \check{\Sigma}(\vr'', t''; \vr', t') \nonumber.
\end{eqnarray}
Subtracting the latter from the former, \ie repeating the step which led to Eq.~(\ref{delt_subtracted}), and Fourier transforming with respect to the relative variables, we see that the self-energy leads to an additional term $\comm{\check{\Sigma}, \check{G}}^{\starp}$ on the right side of Eq.~(\ref{EOM_fourier}). Perfoming the quasiclassical approximations, the Eilenberger equation now gets a term $-\i\comm{\check{\Sigma}, \check{g}}^{\ringp}$ on the right hand side, where $\check{\Sigma} = \check{\Sigma}(\vr, t, \vpfu, \epsilon)$ is the Fourier transformed self-energy with $\vp = \vpf$. Using the unitary transformation in Eq.~(\ref{Unitary_transf}), this leads to a term $-\i\comm{\check{\varsigma}, \check{\G}}^{\ringp}$ in the transformed Eilenberger equation Eq.~(\ref{Eilenberger_transformed}), where $\check{\varsigma} \equiv \hat{U}\check{\Sigma}\hat{U}^\dagger$.

In the diffusive limit we treat the impurity and spin-flip potentials using the self-consistent Born approximation, where the self-energy due to a potential $V(\vr)$ is approximated by  $\check{\Sigma}(\vr, t; \vr', t') = \mean{\check{V}(\vr)\check{G}(\vr, t; \vr', t')\check{V}(\vr')}$.\cite{morten_03,mahan}
For the non-magnetic impurity potential $V_{\mathrm{imp}}$, which we assume to be real, we get the contribution to the self-energy
\begin{eqnarray}
    &&\check{\Sigma}_{\mathrm{imp}}(\vr, t; \vr', t') = \mean{V_{\mathrm{imp}}(\vr)\s^0\check{G}(\vr, t; \vr', t')V_{\mathrm{imp}}(\vr')\s^0} \nonumber \\*
    &&= \frac{1}{\mathcal{V}^2}\Mean{\sum_{i,j}\sum_{\vk, \vk'} e^{\i\vk(\vr-\vr_i)}e^{\i\vk'(\vr'-\vr_j})v_{\mathrm{imp}}(\vk)v_{\mathrm{imp}}(\vk')\check{G}(\vr, t; \vr', t')}\nonumber, \\*
\end{eqnarray}
where we have inserted the Fourier decomposition of the potential from $N_i=n_i \mathcal{V}$ impurities placed randomly in a volume $\mathcal{V}$. The average is done over the impurity positions. Due to the random placement of the impurities, the main contribution from the above expression will come from terms with $i=j$. Performing the sum over $i$ then introduces a factor $N_i\delta_{\vk,-\vk'}$. Since all dependence on impurity positions is gone, averaging is trivial. Letting $\sum_{\vk} \rightarrow \mathcal{V}/(2\pi)^d\int \dif \vk$, where $d$ is the dimension of the Fermi surface, and Fourier transforming with respect to the relative variables, we get
\begin{equation}
    \check{\Sigma}_{\mathrm{imp}}(\vr, t, \vp, \epsilon) = n_i \int\frac{\dif\vq}{(2\pi)^d}|v_\mathrm{imp}(\vp-\vq)|^2 \check{G}(\vr, t, \vq, \epsilon).
\end{equation}
We now use the approximation $\int \dif\vk/(2\pi)^d (\dots) \rightarrow N_0 \int \dif \xi_p \mean{\dots}_\mathrm{F}$\cite{rammer_rmp_86,morten_03}, where $N_0$ is the density of states per spin at the Fermi level, and the definition of the quasiclassical Green's functions in Eq.~(\ref{quasiclassical_integral}) to rewrite the self-energy in terms of the quasiclassical Green's function. Using the unitary transformation in Eq.~(\ref{Unitary_transf}) we arrive at
\begin{equation}
    \check{\varsigma}_{\mathrm{imp}}(\vr, t, \vpfu, \epsilon) = -\i n_i N_0\pi \mean{|v_\mathrm{imp}(\vpf-\vq)|^2 \check{\G}(\vr, t, \vpfu, \epsilon)}_\mathrm{F}.
\end{equation}
Finally, by assuming that the scattering potential is close to isotropic, we arrive at
\begin{equation}
    \check{\varsigma}_{\mathrm{imp}}(\vr, t, \vpfu, \epsilon) = -\frac{\i}{2\tau}\mean{\check{\G}}_\mathrm{F},
\end{equation}
where we have defined the scattering time $\tau^{-1} = 2\pi n_i N_0 \mean{|v_\mathrm{imp}(\vpf-\vq)|^2}_\mathrm{F}$.

We now turn to the spin-flip potential, for which the expression for the self-energy is
\begin{equation}
    \check{\Sigma}_\mathrm{sf}(\vr, t; \vr', t') = \mean{V_\mathrm{sf}(\vr, t)\ve{s}\cdot\paulimat\check{G}(\vr, t; \vr', t')V_\mathrm{sf}(\vr', t')\ve{s}\cdot\paulimat},
\end{equation}
where averaging is done over both impurity locations and impurity spin states $\ve{s}$, both assumed to be random. Following the same procedure as above, we find\cite{morten_03}
\begin{eqnarray}
    &&\check{\Sigma}_\mathrm{sf}(\vr, t; \vr', t) \\*
    &&= \frac{n_\mathrm{sf}}{\mathcal{V}} \sum_\vk e^{\i\vk(\vr-\vr')} |v_{\mathrm{sf}}(\vk)|^2 \frac{S(S+1)}{3}\paulimat\check{G}(\vr,t; \vr', t')\cdot \paulimat,\nonumber
\end{eqnarray}
where $n_\mathrm{sf}$ is the density of spin-flip impurities, and $S$ is the spin quantum number. Fourier transforming, using the quasiclassical approximation and the unitary transformation, we arrive at the self-energy in terms of the transformed quasiclassical Green's functions,
\begin{eqnarray}
    \check{\varsigma}_\mathrm{sf}(\vr, t, \vpfu, \epsilon) &=& -\frac{\i}{3} n_\mathrm{sf}N_0\pi S(S+1) \nonumber\\*
    &\times&\mean{|v_{\mathrm{sf}}(\vpf-\vq)|^2\t^3\otimes\vs\check{\G}(\vr, t, \vpfu, \epsilon)\cdot \t^3\otimes\vs}_\mathrm{F} \nonumber\\*
    &\approx& -\frac{\i}{6\tau_{\mathrm{sf}}}\t^3\otimes \vs \mean{\check{\G}}_F\cdot \t^3\otimes\vs,
\end{eqnarray}
where we have assumed a nearly isotropic scattering potential, and defined the spin-flip scattering time $\tau_{\mathrm{sf}}^{-1} = 2\pi n_\mathrm{sf} N_0S(S+1) \mean{|v_\mathrm{sf}(\vpf-\vq)|^2}_\mathrm{F}$.

\subsection{Check of normalization condition}
The transformed Green's function matrix can in general be parametrized as \cite{bobkova_arxiv, bobkova}
\begin{equation}\label{G_param_general}
    \check{\G} = \breve{\G}'\otimes \s^0 + \breve{\G}'' \otimes\vpfu\cdot\vs + \breve{\G}^{\perp} \otimes\hat{\vp}_{\perp}\cdot\vs + \breve{\G}^3\otimes\s^3,
\end{equation}
where $\hat{\vp}_{\perp} = p_y \hat{x} - p_x \hat{y}$ is a unit vector perpendicular to the direction $\vpfu$ of the Fermi momentum.
As was argued previously, in the quasiclassical limit the dominant terms in $\check{\G}$ should commute with $\vpfu\cdot\vs$. Hence we assume $\breve{\G}^3, \breve{\G}^{\perp} \ll \breve{\G}', \breve{\G}''$. We now insert the above parametrization into the Eilenberger equation, Eq.~(\ref{Eilenberger_transformed}), and neglect all terms containing $\breve{\G}^3$ and $\breve{\G}^{\perp}$ except when multiplied by the Fermi energy. We will use the Born approximation to treat the impurity self-energies. For notational simplicity we also neglect the spin-flip and vector potential terms, which enter in ways similar to the terms considered below. The resulting equation has terms proportional to either of the spin terms in the above parametrization ($\s^0$, $\vpfu\cdot\vs$, $\hat{\vp}_{\perp}\cdot\vs$ and $\s^3$), and can hence be separated into four equations which all must be satisfied separately:
\begin{subequations}
    \begin{eqnarray}
        \vf\vpfu\cdot \nabla \breve{\G}'' &=& \i\comm{\epsilon\t^3 + \frac{\i}{2\tau}\mean{\breve{\G}'}_\mathrm{F} + \i\Delta\t^1, \breve{\G}'}^{\ringp}\nonumber \\*
         &+& \i\comm{\ve{h}\cdot\vpfu\t^3 + \frac{\i}{2\tau}\mean{\breve{\G}''\vpfu}_\mathrm{F}\cdot\vpfu, \breve{\G}''}^{\ringp},\nonumber\\*\label{eil_a}
    \end{eqnarray}
    \begin{eqnarray}
        \vf\vpfu\cdot \nabla \breve{\G}' &=& \i\comm{\epsilon\t^3 + \frac{\i}{2\tau}\mean{\breve{\G}'}_\mathrm{F} + \i\Delta\t^1, \breve{\G}''}^{\ringp} \nonumber \\*
        &+& \i\comm{\ve{h}\cdot\vpfu\t^3 + \frac{\i}{2\tau}\mean{\breve{\G}''\vpfu}_\mathrm{F}\cdot\vpfu, \breve{\G}'}^{\ringp},\nonumber\\*\label{eil_b}
    \end{eqnarray}
    \begin{eqnarray}
        \vf\hat{\vp}_{\perp}\cdot \nabla \breve{\G}' &=& \i\comm{\ve{h}\cdot\hat{\vp}_{\perp}\t^3 + \frac{\i}{2\tau}\mean{\breve{\G}''\vpfu}_\mathrm{F}\cdot\hat{\vp}_{\perp}, \breve{\G}'}^{\ringp} \nonumber \\*
        &+& \anticomm{h_z\t^3, \breve{\G}''}^{\ringp} + 2\epsilon_\mathrm{F} \breve{\G}^3,\label{eil_c}
    \end{eqnarray}
    \begin{eqnarray}
        0 &=& \i\anticomm{\ve{h}\cdot \hat{\vp}_{\perp}\t^3 + \frac{\i}{2\tau}\mean{\breve{\G}''\vpfu}_\mathrm{F}\cdot\hat{\vp}_{\perp}, \breve{\G}''}^{\ringp}  \nonumber \\*
        &+& \comm{h_z\t^3, \breve{\G}'}^{\ringp} + 2\i\epsilon_\mathrm{F} \breve{\G}^{\perp}\label{eil_d}.
    \end{eqnarray}
\end{subequations}
We next multiply Eq.~(\ref{Eilenberger_transformed}) by $\breve{\G}$ from the left and again separately the right, before adding the resulting equations, giving
\begin{eqnarray}
        \frac{\vf}{2}\check{\G}\ringp\anticomm{\nabla\check{\G}, \t^0&\otimes&\vs} + \frac{\vf}{2}\anticomm{\nabla\check{\G}, \t^0\otimes\vs}\ringp\check{\G} \nonumber\\*
        &=& \i\Big[\epsilon\t^3\otimes\s^0+\i\Delta\t^1\otimes\s^0 +\ve{h}\cdot\t^3\otimes\vs \nonumber\\*
        &+& \frac{\i}{2\tau}\mean{\check{\G}}_{\mathrm{F}} 
        -\vf\vpf\cdot\t^0\otimes\vs,\check{\G}\ringp\check{\G}\Big]^{\ringp}.
\end{eqnarray}
Inserting the parametrization in Eq.~(\ref{G_param_general}) into the above equation and separating the different spin terms, keeping terms to the same order as above, we get the equations
\begin{widetext}
\begin{subequations}
    \begin{eqnarray}
        \vf\vpfu\cdot \nabla \anticomm{\breve{\G}'', \breve{\G}'}^{\ringp} &=& \i\Comm{\epsilon\t^3 + \frac{\i}{2\tau}\mean{\breve{\G}'}_\mathrm{F}+\i\Delta\t^1, \breve{\G}'\ringp\breve{\G}' + \breve{\G}''\ringp\breve{\G}''}^{\ringp} 
        + \i\Comm{\ve{h}\cdot\vpfu\t^3+ \frac{\i}{2\tau}\mean{\breve{\G}''\vpfu}_\mathrm{F}\cdot\vpfu, \anticomm{\breve{\G}'', \breve{\G}'}^{\ringp}}^{\ringp},\nonumber\\*\label{norm_a}\\
        \vf\vpfu\cdot \nabla (\breve{\G}'\ringp\breve{\G}' + \breve{\G}''\ringp\breve{\G}'') &=& \i\Comm{\epsilon\t^3 +\frac{\i}{2\tau}\mean{\breve{\G}'}_\mathrm{F} +\i\Delta\t^1,\anticomm{\breve{\G}'', \breve{\G}'}^{\ringp}}^{\ringp} 
        + \i\Comm{\ve{h}\cdot\vpfu\t^3+ \frac{\i}{2\tau}\mean{\breve{\G}''\vpfu}_\mathrm{F}\cdot\vpfu, \breve{\G}'\ringp\breve{\G}' + \breve{\G}''\ringp\breve{\G}''}^{\ringp},\nonumber\\* \label{norm_b}\\
        \vf\hat{\vp}_{\perp} \cdot \nabla (\breve{\G}'\ringp\breve{\G}') &=& \i\Comm{\ve{h}\cdot\hat{\ve{p}}_{\perp}\t^3 + \frac{\i}{2\tau}\mean{\breve{\G}''\vpfu}_\mathrm{F}\cdot\hat{\vp}_{\perp}, \breve{\G}'\ringp\breve{\G}' + \breve{\G}''\ringp\breve{\G}''}^{\ringp} + \Anticomm{h_z\t^3, \anticomm{\breve{\G}'', \breve{\G}'}^{\ringp}}^{\ringp} \nonumber\\*
        &+& 2\epsilon_\mathrm{F}\big(\anticomm{\breve{\G}', \breve{\G}^3}^{\ringp}-\i\comm{\breve{\G}'', \breve{\G}^{\perp}}^{\ringp}\big),\label{norm_c}\\
        \i\vf\hat{\vp}_{\perp}\cdot\Comm{\breve{\G}'', \nabla\breve{\G}'}^{\ringp} &=& \Anticomm{\ve{h}\cdot \hat{\vp}_{\perp}\t^3 + \frac{\i}{2\tau}\mean{\breve{\G}''\vpfu}_\mathrm{F}\cdot\hat{\vp}_{\perp}, \anticomm{\breve{\G}'', \breve{\G}'}^{\ringp}}^{\ringp} - \i\Comm{h_z\t^3, \breve{\G}'\ringp\breve{\G}' + \breve{\G}''\ringp\breve{\G}''}^{\ringp} \nonumber \\*
        &+& 2\epsilon_\mathrm{F}\big(\anticomm{\breve{\G}', \breve{\G}^{\perp}}^{\ringp} + \i\comm{\breve{\G}'', \breve{\G}^{3}}^{\ringp}\big)\label{norm_d}.
    \end{eqnarray}
\end{subequations}
\end{widetext}
Inserting the parametrization Eq.~(\ref{G_param_general}) into the normalization condition in Eq.~(\ref{normalization}) we get, to lowest order, 
\begin{eqnarray}
    \check{\G}\ringp\check{\G} &=& (\breve{\G}'\ringp \breve{\G}' + \breve{\G}''\ringp\breve{\G}'')\otimes \s^0 + \anticomm{\breve{\G}',\breve{\G}''}^{\ringp}\otimes \vpfu\cdot \vs \nonumber \\*&=& 2\t^0\otimes (\s^0 + \vpfu \cdot \vs).
\end{eqnarray}
From this we get the conditions $\anticomm{\breve{\G}'', \breve{\G}'}^{\ringp} = \breve{\G}'\ringp\breve{\G}' + \breve{\G}''\ringp\breve{\G}'' = 2\t^0$, \ie we must assume that $\breve{\G}' = \breve{\G}''$. Hence Eqs.~(\ref{norm_a}) and (\ref{norm_b}) are satisfied by the normalization condition. We can show that the last two of the above equations are also satisfied by using Eqs.~(\ref{eil_c}) and (\ref{eil_d}) together with $\breve{\G}' = \breve{\G}''$, where we write only $\breve{\G}'$ below. For simplicity we define $\breve{A}=\ve{h}\cdot \hat{\vp}_{\perp}\t^3 + \frac{\i}{2\tau}\mean{\breve{\G}''\vpfu}_\mathrm{F}\cdot\hat{\vp}_{\perp}$. For Eq.~(\ref{norm_c}) we get
\begin{widetext}
\begin{eqnarray*}
    \vf\hat{\vp}_{\perp} \cdot \nabla (\breve{\G}'\ringp\breve{\G}') &=& \vf\hat{\vp}_{\perp} \cdot (\breve{\G}' \ringp \nabla \breve{\G}' + \nabla\breve{\G}'\ringp\breve{\G}')\\
    &=& \breve{\G}'\ringp \Big(\i\comm{\breve{A}, \breve{\G}'}^{\ringp} + \anticomm{h_z\t^3, \breve{\G}'}^{\ringp} + 2\epsilon_\mathrm{F} \breve{\G}^3\Big) +  \Big(\i\comm{\breve{A}, \breve{\G}'}^{\ringp} + \anticomm{h_z\t^3, \breve{\G}'}^{\ringp} + 2\epsilon_\mathrm{F} \breve{\G}^3\Big)\ringp\breve{\G}'\\
    &=& \i\comm{\breve{A}, \breve{\G}'\ringp\breve{\G}'}^{\ringp} + 2\epsilon_F\anticomm{\breve{\G}',\breve{\G}^3}^{\ringp} + \breve{\G}'\ringp\anticomm{h_z\t^3, \breve{\G}'}^{\ringp} + \anticomm{h_z\t^3, \breve{\G}'}^{\ringp}\ringp\breve{\G}'\\
    &=& \i\comm{\breve{A}, \breve{\G}'\ringp\breve{\G}'}^{\ringp} + 2\epsilon_F\anticomm{\breve{\G}',\breve{\G}^3}^{\ringp} + \anticomm{h_z\t^3, \breve{\G}'\ringp\breve{\G}'}^{\ringp} \\
    &+& \Big(\i\anticomm{\breve{A}, \breve{\G}'}^{\ringp} + h_z\t^3\ringp\breve{\G}' + 2\i\epsilon_\mathrm{F} \breve{\G}^{\perp}\Big)\ringp \breve{\G}' - \breve{\G}'\ringp \Big(\i\anticomm{\breve{A}, \breve{\G}'}^{\ringp} - \breve{\G}'\ringp h_z\t^3 + 2\i\epsilon_\mathrm{F} \breve{\G}^{\perp}\Big)\\
    &=&\i\Comm{\breve{A}, \breve{\G}'\ringp\breve{\G}' + \breve{\G}'\ringp\breve{\G}'}^{\ringp} + \Anticomm{h_z\t^3, \anticomm{\breve{\G}', \breve{\G}'}^{\ringp}}^{\ringp} + 2\epsilon_\mathrm{F}\big(\anticomm{\breve{\G}', \breve{\G}^3}^{\ringp}-\i\comm{\breve{\G}', \breve{\G}^{\perp}}^{\ringp}\big),
\end{eqnarray*}
where we have used Eq.~(\ref{eil_c}) in the first line, and Eq.~(\ref{eil_d}) in the fourth line. In a similar way we can show that Eq.~(\ref{norm_d}) also is satisfied:
\begin{eqnarray*}
\i\vf\hat{\vp}_{\perp}\cdot\Comm{\breve{\G}', \nabla\breve{\G}'}^{\ringp} &=& \i\vf\hat{\vp}_{\perp}\cdot \breve{\G}'\ringp\nabla\breve{\G}' - \i\vf\hat{\vp}_{\perp}\cdot \nabla\breve{\G}'\ringp\breve{\G}'\\
&=& -\breve{\G}'\ringp\Big(\comm{\breve{A}, \breve{\G}'}^{\ringp} - \i\anticomm{h_z\t^3, \breve{\G}'}^{\ringp} - 2\i\epsilon_\mathrm{F} \breve{\G}^3\Big)+\Big(\comm{\breve{A}, \breve{\G}'}^{\ringp} - \i\anticomm{h_z\t^3, \breve{\G}'}^{\ringp} - 2\i\epsilon_\mathrm{F} \breve{\G}^3\Big)\ringp\breve{\G}'\\
&=& 2\i\epsilon_\mathrm{F}\comm{\breve{\G}', \breve{\G}^3}^{\ringp} + \comm{\breve{A}, \breve{\G}'}^{\ringp}\ringp\breve{\G}' - \breve{\G}'\ringp\comm{\breve{A}, \breve{\G}'}^{\ringp}  - \i \comm{h_z\t^3, \breve{\G}'\ringp\breve{\G}'}^{\ringp}\\
&+& \breve{\G}'\ringp \Big(2\epsilon_\mathrm{F}\breve{\G}^{\perp} + \anticomm{\breve{A}, \breve{\G}'}^{\ringp} + \breve{\G}'\ringp\i h_z\t^3\Big) + \Big(2\epsilon_\mathrm{F}\breve{\G}^{\perp} + \anticomm{\breve{A}, \breve{\G}'}^{\ringp} -\i h_z\t^3\ringp\breve{\G}'\Big)\ringp\breve{\G}' \\
&=&\Anticomm{\breve{A}, \anticomm{\breve{\G}', \breve{\G}'}^{\ringp}}^{\ringp} - \i\Comm{h_z\t^3, \breve{\G}'\ringp\breve{\G}' + \breve{\G}'\ringp\breve{\G}'}^{\ringp} \nonumber + 2\epsilon_\mathrm{F}\big(\anticomm{\breve{\G}', \breve{\G}^{\perp}}^{\ringp} + \i\comm{\breve{\G}', \breve{\G}^{3}}^{\ringp}\big).
\end{eqnarray*}
\end{widetext}
Hence we have shown that the Eilenberger equation is consistent with the normalization condition to lowest order under the assumption that $\breve{\G}' = \breve{\G}''$, which is valid when the exchange energy, spin-flip scattering potential and vector potential is weak compared to the Fermi energy.

\subsection{Usadel equation for $\underline{h}$}
Inserting the parametrization for $\uG^K$ in Eq.~(\ref{h_param}) with $\uh = \uh' + \uh''$ into the Keldysh component of the Usadel equation, Eq.~(\ref{Usadel3D}), we get
\begin{eqnarray}
    2iD\hat{\nabla}\Big(\hat{\nabla}\uh - \uG^R(\hat{\nabla}\uh)\uG^A &+& (\uG^R\hat{\nabla}\uG^R)\uh - \uh(\uG^A\hat{\nabla}\uG^A)\Big) \nonumber\\*
    &=& \comm{\epsilon\t^3,\uG^R\uh-\uh\uG^A},
\end{eqnarray}
where we for simplicity have kept only the first term on the right hand side of the Usadel equation. Inserting the parametrization $\uh = h_L\t^0+h_T\t^3$,\cite{belzig_sam_99,chandrasekhar_arxiv_04} we proceed by multiplying with the identity and $\t^3$ and taking the trace.\cite{chandrasekhar_arxiv_04} This gives the two equations
\begin{eqnarray}
            0 &=& \nabla\cdot\Big\{\nabla h_L\Tr\{\t^0-\uG^R\uG^A\} - \nabla h_T\Tr\{\uG^R\t^3\uG^A\}\nonumber\\*
             &+& h_T\Tr\{\t^3\uG^R\hat{\nabla}\uG^R-\t^3\uG^A\hat{\nabla}\uG^A\}\Big\},\label{ddh_L_exact}
\end{eqnarray}
and
\begin{eqnarray}
        0 &=& \nabla\cdot\Big\{\nabla h_T\Tr\{\t^0-\uG^R\t^3\uG^A\t^3\} - \nabla h_L\Tr\{\uG^R\uG^A\t^3\} \nonumber\\*
        &+& h_L\Tr\{\t^3\uG^R\hat{\nabla}\uG^R-\t^3\uG^A\hat{\nabla}\uG^A\}\Big\}.\label{ddh_T_exact}
\end{eqnarray}
In situations where the last two traces in both the above equations are zero (for instance, the third term corresponds to a supercurrent and is absent in S/N or S/F bilayers), we get two decoupled second order equations for $h_L$ and $h_T$,
\begin{eqnarray}
    \Tr\{\t^0 &-& \uG^R\uG^A\}\nabla^2h_L\\*\label{ddh_L}
    &=& (\nabla h_L)\Tr\{(\nabla\uG^R)\uG^A + \uG^R(\nabla\uG^A)\},\nonumber\\
    \nonumber\\
    \Tr\{\t^0 &-& \uG^R\t^3\uG^A\t^3\}\nabla^2h_T \label{ddh_T}\\*
    &=& (\nabla h_T)\Tr\{(\nabla\uG^R)\t^3\uG^A\t^3 + \uG^R\t^3(\nabla\uG^A)\t^3\}.\nonumber
\end{eqnarray}
In this case the expression for the charge current simplifies to
\begin{equation}
    \bm{j}_q = N_0eD \int\dif\epsilon ~\nabla h_T\Tr\{\t^0-\uG^R\t^3\uG^A\t^3\}\label{j_hT}.
\end{equation}

\end{document}